\documentclass[12pt]{article}
\pdfoutput=1
\usepackage[utf8]{inputenc}
\usepackage{setspace}
\usepackage[margin=1in]{geometry}
\usepackage{appendix}
\usepackage{hyperref}
\usepackage{graphicx}
\usepackage{tabularx}
\usepackage{amsmath}
\usepackage{indentfirst}
\usepackage{float}
\usepackage{lscape}
\usepackage{textcomp}

\usepackage{natbib}
\setcitestyle{authoryear,open={(},close={)}}

\title{Payroll Tax Incidence: Evidence from Unemployment Insurance}

\date{March 2023}
\author{Audrey Guo\thanks{Santa Clara University. Contact: aguo@scu.edu. I would like to thank Katarzyna Bilicka, Maria Fitzpatrick, Tatiana Homonoff, John Ifcher, Kris Mitchener, and seminar participants at Georgetown Law, UC Davis, Santa Clara University, IIPF Annual Congress, and All-California Labor Conference for helpful comments. Any views expressed are those of the authors and not those of the U.S. Census Bureau. The Census Bureau's Disclosure Review Board and Disclosure Avoidance Officers have reviewed this information product for unauthorized disclosure of confidential information and have approved the disclosure avoidance practices applied to this release. This research was performed at a Federal Statistical Research Data Center under FSRDC Project Number 1632. (CBDRB-FY22-P1632-R9305-R9674) This research uses data from the Census Bureau's Longitudinal Employer Household Dynamics Program, which was partially supported by the following National Science Foundation Grants SES-9978093, SES-0339191 and ITR-0427889; National Institute on Aging Grant AG018854; and grants from the Alfred P. Sloan Foundation.}}

\begin{document}

\maketitle

\doublespacing

\begin{abstract}
Economic models assume that payroll tax burdens fall fully on workers, but where does tax incidence fall when taxes are firm-specific and time-varying? Unemployment insurance in the United States has the key feature of varying both across employers and over time, creating the potential for labor demand responses if tax costs cannot be fully passed on to worker wages. Using state policy changes and matched employer-employee job spells from the LEHD, I study how employment and earnings respond to payroll tax increases for highly exposed employers. I find significant drops in employment growth driven by lower hiring, and minimal evidence of pass-through to earnings. The negative employment effects are strongest for young and low-earning workers. 

\end{abstract}
\newpage
\section{Introduction}
Employer taxes on labor earnings are used to fund a variety of social insurance programs, and policymakers frequently view payroll tax cuts as a way to stimulate employment. In the theory of tax incidence, the resulting employment and wage change depends on the elasticities of labor demand and labor supply; because labor demand is assumed to be much more elastic than labor supply (especially in the long run), most models assume payroll tax costs are born entirely by workers, resulting in little to no effect on employment.\footnote{As evidenced by early estimates from \cite{brittain} and \cite{gruber97}.} However, there may be imperfect pass-through in the presence of downward wage rigidities, leading to a decline in employment rather than wages.\footnote{Using data from a large payroll processing firm, \cite{GHY} find very low rates of wage cuts, providing strong evidence for downward nominal wage rigidity. Using Census Bureau data, \cite{murray} estimates that downward nominal wage rigidity accounted for at least 23\% of excess job destruction during the financial crisis of 2008.} \cite{hamermesh} points out that if the entire burden of a payroll tax could be shifted onto wages, employers would not so vehemently oppose tax increases. Furthermore, payroll tax changes that only affect a subset of firms may prevent employers from fully passing costs through to wages.\footnote{Likewise, \cite{saez2012} and \cite{saez2019} found that payroll tax changes only affecting a subset of workers within a firm also failed to result in full pass-through.}
% and existing empirical estimates range from zero to full incidence on workers (\cite{brittain}, \cite{hamermesh}, \cite{gruberkrueger}, \cite{gruber97})

In this paper, I exploit state Unemployment Insurance (UI) tax increases to estimate payroll tax incidence in the United States. The third largest payroll tax in the U.S., UI is an employer-specific and time-varying tax that is administered at the state level. Employer tax rates are an increasing function of the amount of UI benefits paid out to previous employees, and in 2010 and 2011, state-legislated UI tax increases raised the payroll tax costs for highly exposed employers. I use administrative data from the US Census Bureau's Longitudinal Employer-Household Dynamics (LEHD) to study the impact of these state tax increases on worker earnings and employment growth.

I identify a subset of highly exposed employers, defined as those with a history of previous layoffs, and use a difference-in-differences research design to compare the outcomes of employers based in treatment versus control states. I estimate minimal pass-through of state tax increases to earnings, and instead find negative impacts on employment growth. For each \$100 increase in payroll taxes per worker, employment growth declined by 0.43 percentage points over the course of the year. Event study estimates found the largest effects in the 2nd and 3rd year after the initial policy changes, which suggest that UI tax increases hindered the economic recovery after the Great Recession. These results are further confirmed using a triple-difference research design that compares high exposure and low exposure employers within the same state, to control for potential differences in state economic conditions in the wake of the Great Recession.

I test whether this drop in employment growth is driven by increased separations or reduced hiring, and find clear evidence of the latter. The quarterly share of new hires fell by up to 0.2 percentage points for each \$100 tax increase, while the effect on the separation share was small and statistically insignificant. These findings align with firm tax incentives, as each additional new hire would incur another instance of UI tax; on the other hand, additional separations provide no UI tax savings, since low UI tax bases cause most earnings to exceed the tax base by the end of the first quarter. Heterogeneity analysis shows that the negative employment effects are stronger for young (age under 25) and low-earning workers. Low-earning workers also experienced larger drops to earnings growth than the average worker, but still well below full pass-through. Additionally, single-establishment firms were more likely to reduce hiring than multi-establishment firms, suggesting that cash constraints may also play an important role.

The main contribution of this paper is providing a U.S. context to study payroll tax incidence, as the lack of variation in federal payroll taxes has caused US-based studies to be largely absent from the recent literature. Given differences in labor market institutions, the estimates of tax incidence in one country may not generalize in the same way to another. State-level UI tax increases provide a setting similar to \cite{kuetal}, who found that regional payroll tax increases in Norway lowered employment, because firms were only partially able to pass the costs on to workers. \cite{saez2021} finds that the positive employment effects from a young workers payroll tax cut persisted even after workers aged out of eligibility. Likewise, I estimate a gradual and sustained decrease in employment growth due to payroll tax increases. \cite{benzarti2} found that a payroll tax cut in Finland helped treated firms weather the Great Recession despite having minimal impact on employment beforehand, suggesting a cyclical component to firm responses. My findings of a negative employment response to tax increases following the Great Recession confirms the symmetry of this effect.

Another contribution is the estimation of heterogeneous effects across worker and employer characteristics. \cite{benzarti} found that firms with greater payroll taxes substituted away from low-skilled and manual workers. I find greater drops in labor demand for low-earning and young workers, but not for those with less education. My finding that single-establishment firms were especially likely to decrease hiring in response to higher UI taxes, is consistent with the finding in \cite{lobel} that small firms were more responsive to payroll tax cuts. \cite{hungarypayroll} estimate heterogeneity in payroll tax incidence by firm productivity, finding that lower productivity firms experienced employment responses while higher productivity firms passed tax cuts on to workers. While I do not observe firm productivity, my sample of high exposure employers likely overlaps with lower productivity firms, as they suffered layoffs prior to the tax increases.

Finally, my paper contributes to the small literature on UI tax incidence by revisiting the question using recent data and exogenous policy changes from multiple states. Early studies by \cite{andersonandmeyer1997} and \cite{andersonandmeyer2000} found evidence of pass-through for industry-specific UI taxes but not employer-specific rates, although the variation in employer-specific rates included both high and low exposure employers together. My findings also complement the results in \cite{johnston}, which found little effect of annual changes in employer-specific rates on worker earnings, but large impacts on hiring.

The next section provides institutional details on US Unemployment Insurance. Section 3 describes the research design, and Section 4 discusses the data and construction of my analysis sample. Section 5 presents the main employer-level results, and Section 6 presents a worker-level analysis focusing on low-earning workers. Section 7 concludes.

\section{Institutional Background}
Unemployment insurance in the U.S. is funded solely through payroll taxes on employers.\footnote{With the exception of New Jersey, which also charges workers a uniform payroll tax of 0.5\%.} Unlike Social Security or Medicare taxes, which are uniform across all employers, UI is experience-rated and therefore employer-specific. UI tax revenues reached a height of 1 percent of total wages in 2012, but this masks considerable heterogeneity across employers. Employers that lay off workers who then claim UI benefits are assigned a higher future tax rate - up to a statutory maximum.\footnote{\cite{guojohnston} provides a brief history of U.S. unemployment insurance, which was modeled after the experience rating in workers' compensation. Since its inception, policymakers have recognized the tradeoff between disincentivizing layoffs, and the cyclical tax burdens that dynamic experience rating would create.} These employer-specific rates are calculated annually, and Appendix \autoref{sample_sched} illustrates a sample tax schedule from the state of Florida, where the maximum tax rate is capped at 5.4\%. Appendix \autoref{cycles} plots total UI contributions over time as a percentage of total wages, and reveals the cyclical pattern of UI tax collections, where tax rates are highest in the years following recessions. Due to this time-varying nature of UI tax rates, businesses may have more difficulty passing the costs on to their workers, and changes in UI tax liabilities have the potential to influence employment as well as earnings. 

Another unique aspect of UI is the low tax base in most states. Throughout the last half-century, taxable wages have not kept up with inflation, leading to an errosion of the tax base. In 2009, the year before the first UI tax increases, the median UI tax base was only \$10,000 -- with the median state's taxable wages covering only 28\% of total payroll. This low tax base causes UI to become a lump-sum tax on workers, independent of earnings. Thus UI taxation is regressive in the sense that seasonal, part-time, and low-wage workers are taxed at a higher effective tax rate (as a share of earnings), and these types of workers become relatively more costly for a business facing high UI tax rates. \cite{huang} finds that higher state taxable wage bases increases teenage employment rates, and \cite{dgjpandp} finds that higher tax bases increase labor demand for low-wage part-time workers.

There is substantial variation in UI tax schedules across states, and a tax increase in one state raises payroll costs for incumbent employers, without affecting tax regimes in any other. Many state UI programs are underfunded, leading them to deplete their UI trust funds during recessions. During the Great Recession, 36 states depleted their UI trust funds and were forced to take out loans from the federal government; this occurred again for 18 states during the height of the COVID-19 pandemic. While many states recognized the need to update their UI tax schedules to replenish their trust funds following the Great Recession, the timing and the size of the legislated tax increases varied greatly. At the same time, there were proposals to increase the federally mandated UI tax base (of \$7,000), which would have given states a way to obtain more tax revenue without needing to legislate their own UI tax policy. California, for example, did not legislate any UI tax increases despite a UI trust fund debt of over \$10 billion.

\section{Research Design}
\begin{figure}[H]
\begin{minipage}{\linewidth}
    \caption{Maximum UI Tax Increase from 2009 to 2011}
       \label{taxchange}
    \centering
\includegraphics[width=\linewidth]{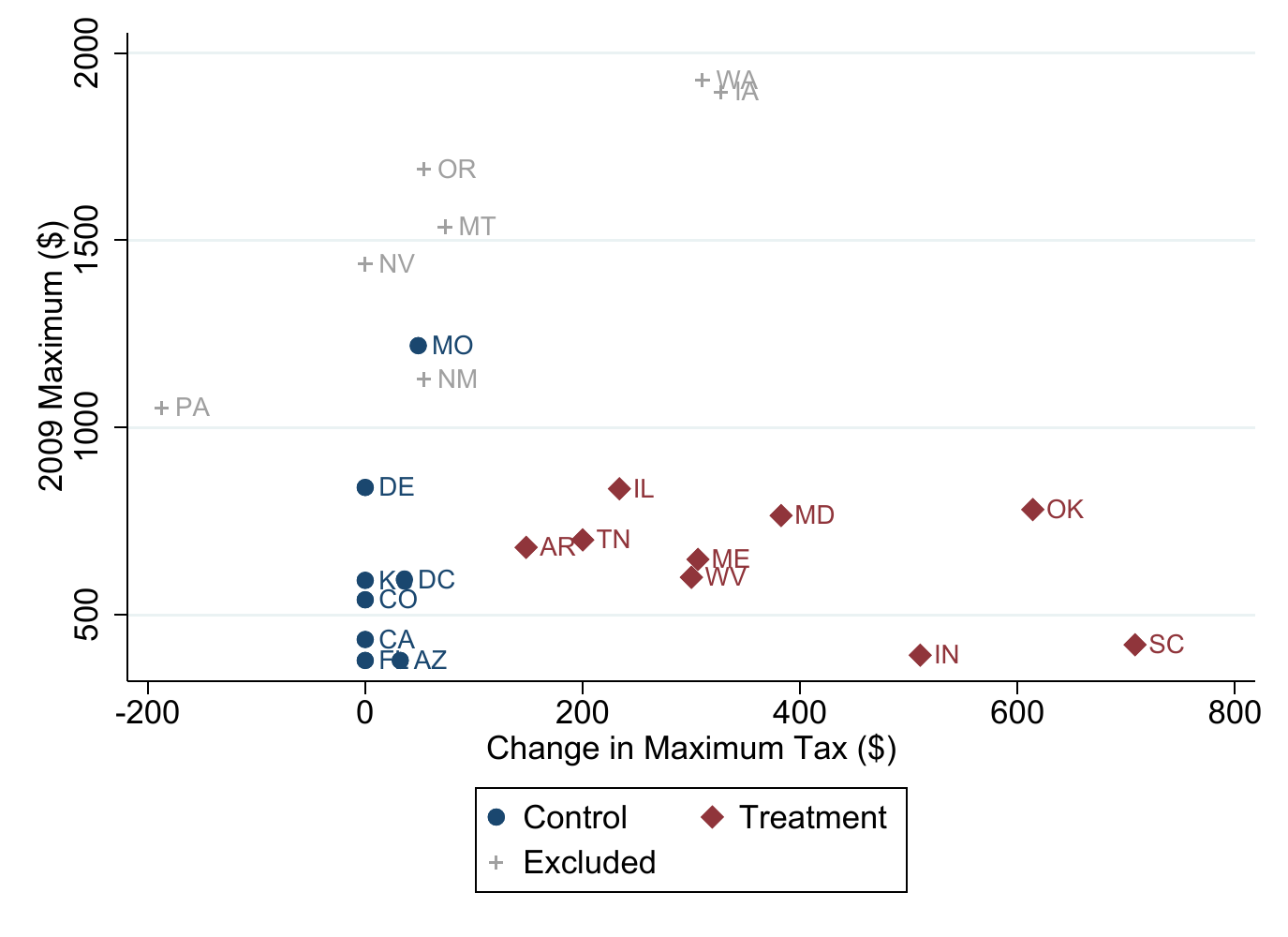}
\end{minipage}
\singlespacing
Source: \textit{Significant Measures of State UI Tax Systems}. Maximum tax calculated as taxable wage base multiplied by maximum tax rate.
\end{figure}

In 2010 and 2011, a total of sixteen states implemented UI tax increases in order to replenish their trust funds. Coupled with the mechanical increase in employer-specific tax rates due to experience rating, these tax hikes could amount to a 2-5\% increase in effective tax rates. For my LEHD research sample of 23 states and District of Columbia, \autoref{taxchange} graphs the distribution of maximum tax increases (tax base * maximum rate) from 2009 to 2011. Among this subsample, I exclude 7 states from my analysis due to basline differences in UI tax regimes. Six of these excluded states automatically index UI tax bases to wage growth, resulting in incremental and predictable tax increases every year throughout my sample period (and thus greater tax costs at baseline than the treatment states). I also exclude Pennsylvania, which experienced a UI tax \textit{decrease} in 2011-12. 

\begin{figure}[H]
\begin{minipage}{\linewidth}
    \caption{UI Tax Policy Variation}
    \centering
    \includegraphics[width=\linewidth]{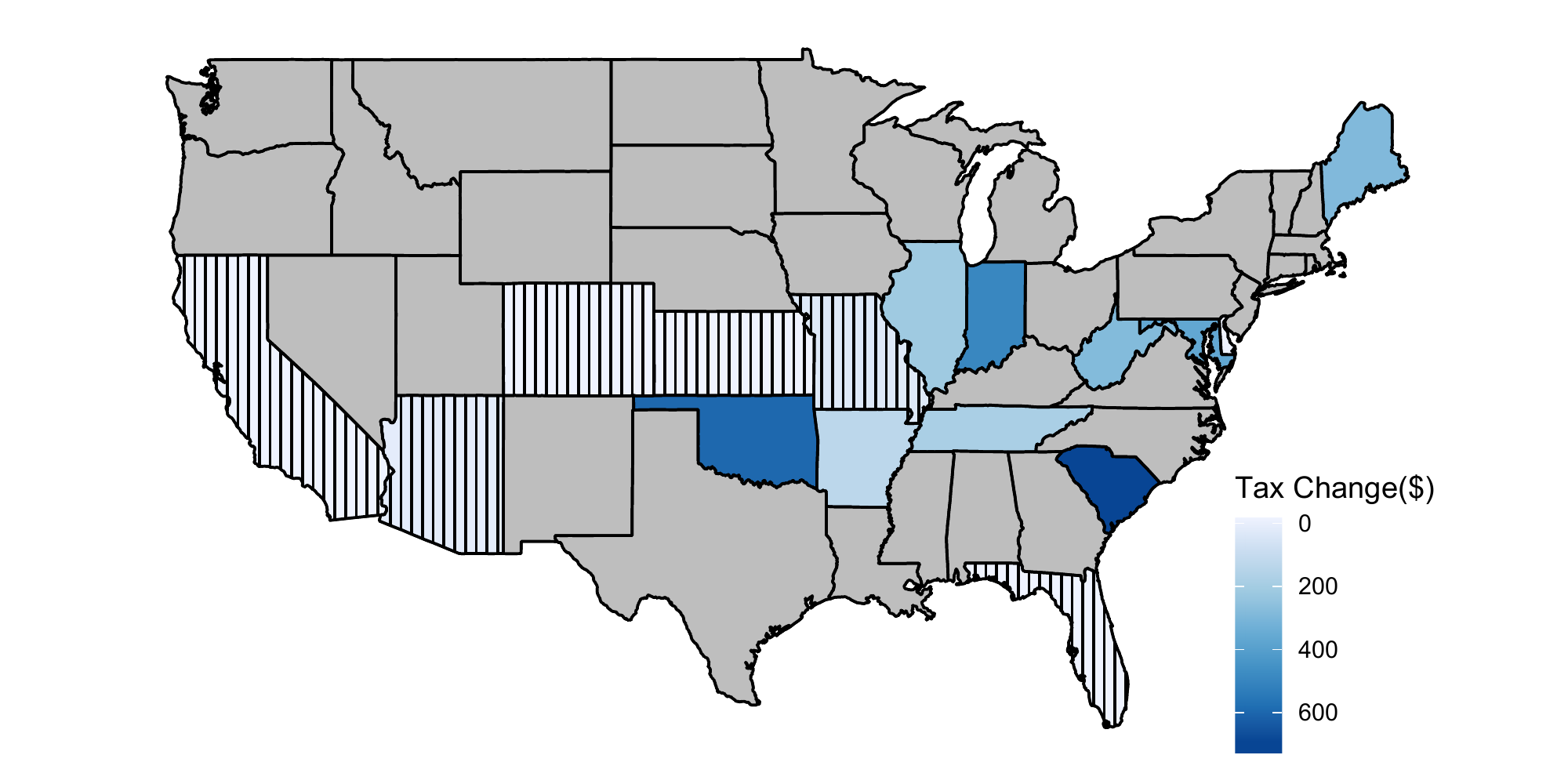}
   \label{lehdmap}
\end{minipage}
\singlespacing
Source: \textit{Significant Measures of State UI Tax Systems}. The 7 treatment states are shaded in blue corresponding to the tax change they experienced, and the 8 control states are shaded with vertical lines (and also include Delaware and District of Columbia). States in grey are excluded.
\end{figure}

From the remaining states, I identify 9 treatment states whose statutory maximum UI taxes rose by more than \$100 from 2009 to 2011, leaving a remainder of 8 control states with minimal tax changes during this window. From 2009 to 2011, treatment states experienced per-capita maximum UI tax increases ranging from just under \$200 to over \$600, and the average increase was \$378. In percentage terms, maximums increased by 67\% on average, and the largest tax increase was experienced by South Carolina (in both level and percentage terms).\footnote{Appendix \autoref{statebystate} plots maximum and average UI taxes over time for each control state, and appendix \autoref{statebystate2} and  \autoref{statebystate3} plots them for treatment and excluded states, respectively.}  \autoref{lehdmap} plots the locations of these 17 LEHD states, and the change in maximum UI taxes from 2009 to 2011.

In order to focus on employers for whom these state UI tax increases would be binding, I create a sample of high exposure employers that are likely to be close to the maximum tax rate following the Great Recession. I first estimate event study regressions to test for differential pre-trends in treatment states relative to control states. I then estimate pooled difference-in-differences regressions to estimate a continuous treatment effect from UI tax increases. To establish a consistent sample for both regression specifications, I restrict to the period 2008 to 2013, to allow for at least two years of pre-change data and three years of post-change data.

Estimating an event study requires defining a single starting date for each state tax increase. While some state increases occurred all at once, other states legislated a more gradual increase from 2010 to 2012. The pooled difference-in-differences estimation will incorporate this variation in magnitude and timing, but for the event study I simply define event time relative to the quarter before the initial tax increase occurs ($event = 0$ in either 2009:Q4 or 2010:Q4 for treatment states, and event time always equals zero for control states). I then estimate the following regression specification:

\begin{equation}
      Y_{fst} = \alpha_{f} + \sum_{k=-8, k\neq -1}^{13}\beta_{k}(event = k) + NAICS*\delta_{t} + \gamma minwage_{st} + \epsilon_{fst}     
      \label{eq1}
\end{equation}

Here $f$ indexes employer (SEIN), $s$ indexes state, $t$ indexes year-quarter, and $k$ indexes the quarter relative to the policy change. I group all quarters more than two years before the policy change together into one estimate $k=-8$, and all quarters more than three years after the policy change together into $k=13$. To allow for potential anticipation effects, the treatment effects are estimated relative to the outcome two quarters prior to the tax change (ie: if a state increased taxes in 2010, the baseline would be Q3 of 2009, and if a state increased taxes in 2011, the baseline would be Q3 of 2010). Thus the excluded quarter is $k=-1$, and the first quarter of the tax change is labeled $k=1$.\footnote{States with 2010 increases are Arkansas, Maine, Maryland, Tennessee, and West Virginia; States with 2011 increases are Illinois, Indiana, Oklahoma, and South Carolina.} Coefficients $\beta_{-8}$ to $\beta_{-2}$ test for differential pre-trends prior to the policy changes. Fixed effects are included for the SEIN, as well as the year-quarter interacted with 2-digit NAICS sector, as some sectors such as construction were disproportionately impacted by the Great Recession. The primary outcomes of interest include mean log quarterly earnings for stable employees (excludes hires and separations), as well as year-over-year employment growth and hiring rates. 

I also estimate a pooled difference-in-differences regression with a continuous treatment variable, $Taxchange_{st}$, defined as the dollar change in maximum UI tax relative to 2009 (and defined to equal zero in years 2008 and 2009). The baseline regression specification is as follows:
\begin{equation}
       Y_{fst} = \alpha_{f} + \sum_{q=1}^{4}\beta_{q}*Taxchange_{st} + NAICS*\delta_{t}  + \gamma minwage_{st} + \epsilon_{fst} 
       \label{eq2}
\end{equation}
 
 Here $f$ indexes employer, $s$ indexes state, and $t$ indexes year-quarter. I estimate four coefficients of interest $\beta_1$ to $\beta_4$ to allow for differential responses by calendar quarter, because UI tax burdens are largest in Q1. The identifying assumption is that in the absence of UI tax policy changes, employers in the same 2-digit NAICS sector in treatment states and control states would have evolved similarly in the years following the Great Recession.

 It is worth noting that after the Great Recession, there was also a significant federal extension of UI benefit duration beyond the typical 26 weeks, through both the Emergency Unemployment Compensation (EUC) program and the Extended Benefits (EB) program. Benefit duration reached a potential maximum of 99 weeks in 2010-2011, before dropping in 2012 and coming abruptly to an end at the close of 2013. \cite{farbervalletta} and \cite{frv_pandp} found little impact of these benefit extensions on job finding, as they primarily caused fewer workers to leave the labor force and led to long-term unemployment. Thus, I assume the extension of UI benefits does not differentially impact worker labor supply during my analysis period.

\section{Data}
This project uses the Census Bureau's Longitudinal Employer-Household Dynamics (LEHD), an administrative employer-employee matched dataset of quarterly earnings, directly sourced from state UI records. Each employer in the LEHD is assigned a state EIN number (SEIN), and this will be my firm definition. A SEIN is also the level at which UI tax rates are assigned to employers. It can encompass multiple establishments within a state, but no employment outside the state. And just as large firms may have multiple EINs associated with their business, these firms may also own multiple SEINs within a given state.

To construct my sample, I include employers from 9 treatment states and 8 control states, covering the time period from 2008 to 2013 (allowing for an unbalanced panel). I also drop the Retail Trade and Accommodation \& Food Services sectors (NAICS 44-45 and NAICS 72). These two sectors experience high turnover and low UI take-up rates, such that it is very difficult to classify employers into high versus low experience rating. The exclusion of these sectors should not affect the generalizability of my results, as they rank among the sectors with the lowest average UI tax costs, just above Healthcare and Education (\cite{guojohnston}).

The first outcome of interest is average quarterly earnings, either calculated for stable workers (which excludes hires and separations) or for all workers. Since no information on hours is collected, incidence on earnings may also reflect changes in hours. In additional analyses in Section 6, I identify low-earning workers in the LEHD to separately estimate the impacts on these workers with the highest relative UI tax burdens. To study employment outcomes, I count hires and separations based on the beginning and end of employment spells, and divide by employment to define quarterly shares of new hires and separations. I also measure quarterly employment growth using DHS growth rates.

\subsection{Calculating UI Tax Exposure}
State UI tax increases affected either the maximum tax rate or tax base, both of which require a sufficiently high experience rating to be binding. Because I do not directly observe employers' payroll tax rates in the LEHD, I impute exposure to UI tax increases based on separation history. Employers who never separate workers into non-employment are inferred to have a low UI tax rate (low exposure), and employers who separate a significant share of workers into non-employment will be inferred to have a tax rate close to the state maximum (high exposure).

It is not layoffs themselves that impact employer-specific tax rates, but rather the amount of UI benefits claimed by laid off workers. In order to identify separations that could be eligible for UI benefits, I first limit the scope of consideration to workers with at least 2 consecutive quarters of earnings, and high quarter earnings of at least \$1500 (minimum requirement to be eligible for UI benefits in most states). I also ignore workers younger than age 18 (to abstract from job separations driven by schooling) or older than age 60 (to abstract from retirement timing). I then impute a layoff as a job separation in quarter $t$ that results in at least one quarter of zero earnings from quarters $t+1$ to $t+3$. Separations that were the result of moves to a different state will not be captured in the LEHD data, so these could also be miscoded as layoffs. I also drop employers with less than 20 or greater than 500 workers at baseline (2009:Q3), to minimize measurement error in this imputation.

Then for each SEIN, I aggregate the number of imputed layoffs over two separate pre-periods, and use these two cumulative shares to define inclusion in my analysis sample as a high exposure firm. The first pre-period runs from 2006:Q1 to 2007:Q4, and the second runs from 2008:Q1 to 2009:Q2 (the Great Recession began in December 2007 and ended in June 2009).\footnote{I use the three-year period before the first state tax increases in order to best capture layoffs that could lead to potential UI tax increases. Layoffs that occurred prior to 2006 are unlikely to impact tax rates in 2010 and later, because most firms will have already finished experiencing tax increases. For example, layoffs during the 2008-09 Great Recession caused tax rates to peak in 2011-12.} In both treatment and control states, I restrict the sample to SEINs that in each of the two pre-periods experienced total layoffs of 33-100\% of their 2009:Q3 baseline employment. 

I also reserve a subset of SEINs with imputed layoffs of less than 15\% from 2006:Q1-2007:Q4, and less than 10\% from 2008:Q1-2009:Q2. Given the likelihood of overestimating layoffs, an SEIN with an imputed layoff rate less than 10\% very likely experienced zero or very little UI benefit claims. This set of employers, which are expected to be the least affected by UI tax increases due to their distance from the maximum tax rate, will act as a ``placebo" or low exposure sample in triple-difference analyses. 

Separation into non-employment is still an imperfect predictor of eventual experience rating, because of imperfect take-up rates and substantial variation in UI duration. Thus I view this exercise as accepting there will be Type I error, in favor of minimizing Type II error. My results are also robust to using alternative thresholds to define this sample of high exposure employers. In new work using administrative data from Washington, \cite{wash_ui} find that the take-up of UI among workers plays as important a role as the separation rate in determining a given employer’s future UI tax rate. Thus the measurement error introduced by inadvertently including some employers with low UI tax exposure may attenuate my estimates. 

\subsection{Summary Statistics}
\autoref{summstats} reports summary statistics for my sample of firms in treatment and control states at baseline in the third quarter of 2009. They are very similar in size, with average employment of roughly 67 workers. Average layoff histories also match extremely well, with the average employer having experienced cumulative layoffs in 2006-07 that amounted to 57\% of their 2009:Q3 employment, and 50\% for the period of 2008:Q1 to 2009:Q2. One area of mismatch, however, is annual earnings, with control states paying higher wages than treatment states. This is likely due to state-level differences in cost-of-living and minimum wages. Federal mininum wage increases occurred in July of 2007, 2008, and 2009, and treatment states were more likely to have faced binding minimum wage increases. Thus I also include a control for the state minimum wage in all of my regression specifications. This table also shows that the construction sector is overrepresented among high-exposure firms; nationally the industry only accounts for 10\% of establishments, while in my analysis sample construction employers make up over 20\%.

\section{Results}
I first establish that increases to the statutory maximum UI tax translated into increases in actual taxes paid. In the absence of employer-specific tax rates, I estimate an event study regression using aggregate data from the public-use Quarterly Census of Employment and Wages (QCEW). The observation level is a 4-digit industry, state, and quarter, and effective UI tax rates are calculated by dividing quarterly UI contributions by taxable quarterly payroll. 

\begin{figure}[H]
\begin{minipage}{0.9\linewidth}
    \caption{Effective UI Tax Rates (2008-2013)}
       \label{erate2}
    \centering
\includegraphics[width=\linewidth]{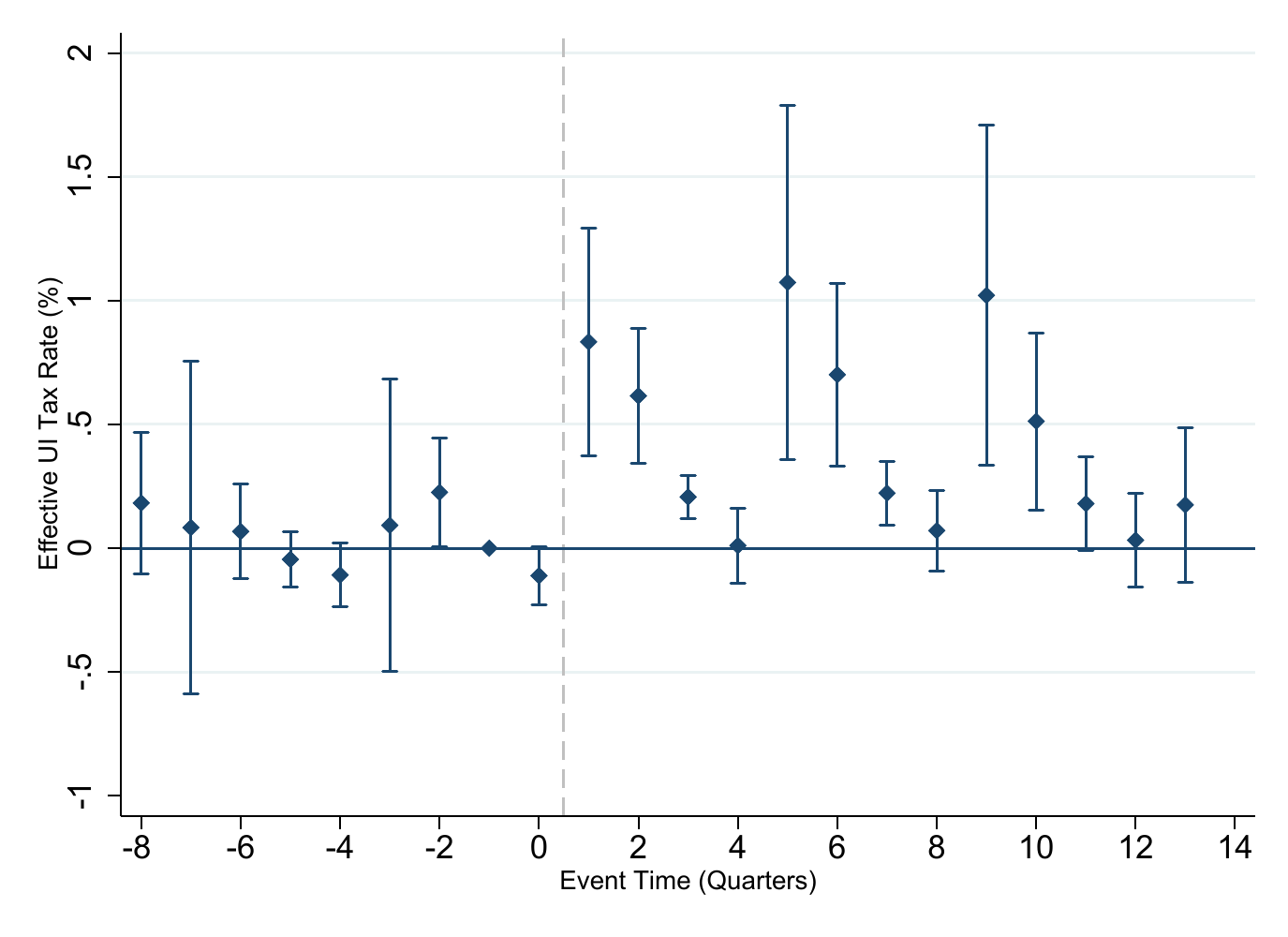}
\end{minipage}
\singlespacing
Source: \textit{Quarterly Census of Employment and Wages}. (N = 4,062) Sample limited to ten 4-digit construction industries and years 2008 to 2013 (cells too small to meet disclosure requirements are missing). Effective tax rates calculated by dividing quarterly UI contributions by quarterly payroll. Event study estimated with state-industry and year-quarter fixed effects, and weighted by industry employment. Error bars denote 95\% CI for standard errors clustered at state level.
\end{figure}

Using construction industries as a proxy for high exposure employers (as they have the highest average tax costs), \autoref{erate2} shows that the policy changes in treatment states led to an increase in UI tax costs, predominantly in the first two quarters of the calendar year (Q1 coincides with event times -7, -3, 1, 5, and 9). On average, Q1 tax rates in treatment states increased by 0.83, 1.1, and 1.0 percentage points over the three years following a policy change, relative to control states. Estimates from a difference-in-differences specification in Appendix \autoref{erateDD} find that each \$100 increase in maximum UI taxes increased effective UI tax rates by 0.16 percentage points in the first quarter of the year. However, this average effect at the industry level masks substantial heterogeneity in effective tax rates that occurs at the worker level, especially for workers that earn below the taxable wage base. This will be explored further in section 6.

\subsection{Event Study}
\begin{figure}[H]
\begin{minipage}{\linewidth}
    \caption{Event Study of Log Quarterly Earnings (2008-2013)}
    \vspace{2mm}
       \label{earn}
    \centering
A: Baseline Estimates \\
\includegraphics[width=0.9\linewidth]{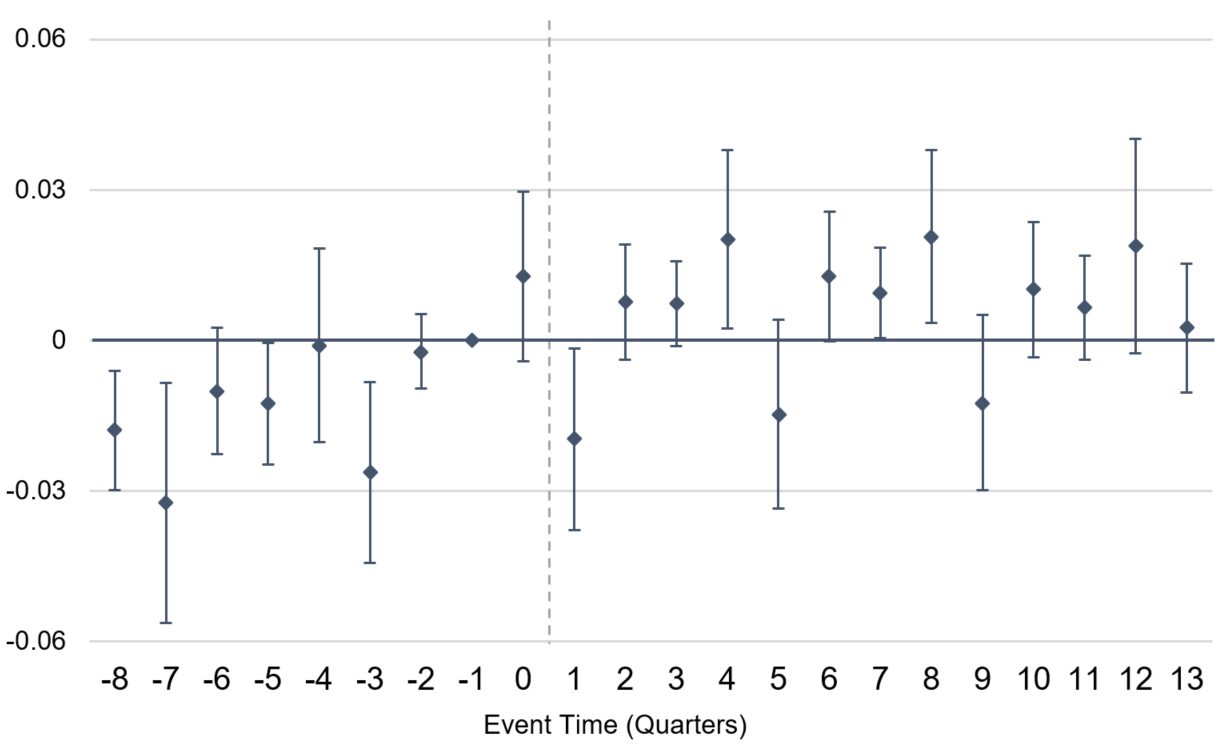} \\
B: Controlling for State-by-Calendar Quarter \\
\includegraphics[width=0.9\linewidth]{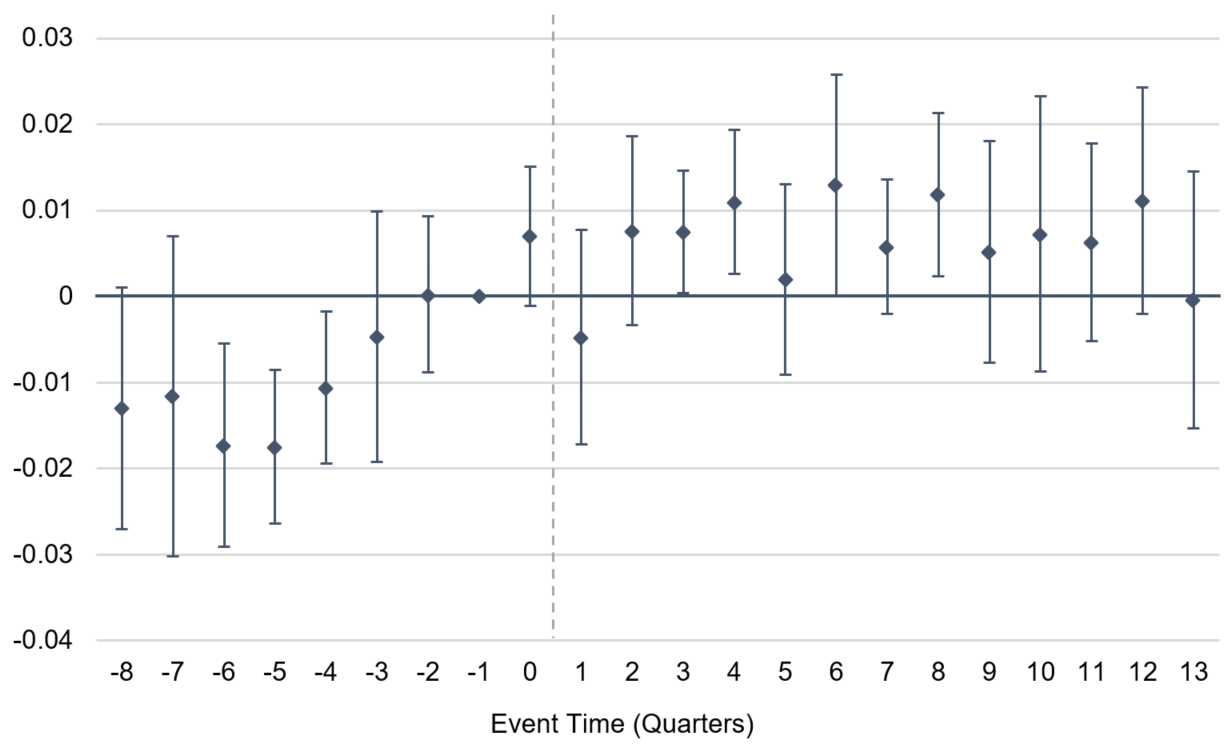}
\end{minipage}
\singlespacing
N = 651,000. Outcome variable is the mean quarterly earnings of stable workers (employed in both t-1 and t+1). Also includes a control for state minimum wage, and includes SEIN and Year-Quarter-N2 fixed effects. Panel B also includes state-by-calendar quarter fixed effects. Error bars denote 95\% CI for robust standard errors clustered at state level.
\end{figure}

To study the impact on earnings, \autoref{earn} plots event study estimates from \autoref{eq1} using log quarterly earnings as the outcome. Log earnings are calculated by taking the firm-level average of log quarterly earnings for all stable employees continuously employed from quarter t-1 to t+1. By excluding new hires and separations, this ignores any potential drops in earnings due to incomplete job spells. The baseline estimates in Panel A show that even after controlling for state minimum wages, employers in treatment states paid lower wages than those in control states prior to the policy change, and had a pattern of lower first quarter earnings relative to the rest of the year. This persistent seasonality in earnings is likely driven by regional differences in climate. Two of the largest control states, California and Florida, are known to have very mild winters and thus less seasonal employment in high exposure industries such as construction. Panel B shows that including state-by-calender quarter fixed effects helps to smooth out these seasonal patterns. After controlling for state-specific seasonality, we still observe small drops in average earnings in the first quarter (when tax burdens are greatest), but the magnitudes are very small and not statistically significant (implying at most a 1.2\% drop in Q1 earnings). This therefore suggests that UI tax increases have a minimal impact on workers' average quarterly earnings.

\begin{figure}[H]
\begin{minipage}{\linewidth}
    \caption{Event Study of Year-Over-Year Employment Growth (2008-2013)}
        \vspace{2mm}
           \label{growthyoy}
    \centering
A: Baseline Estimates \\
\includegraphics[width=0.9\linewidth]{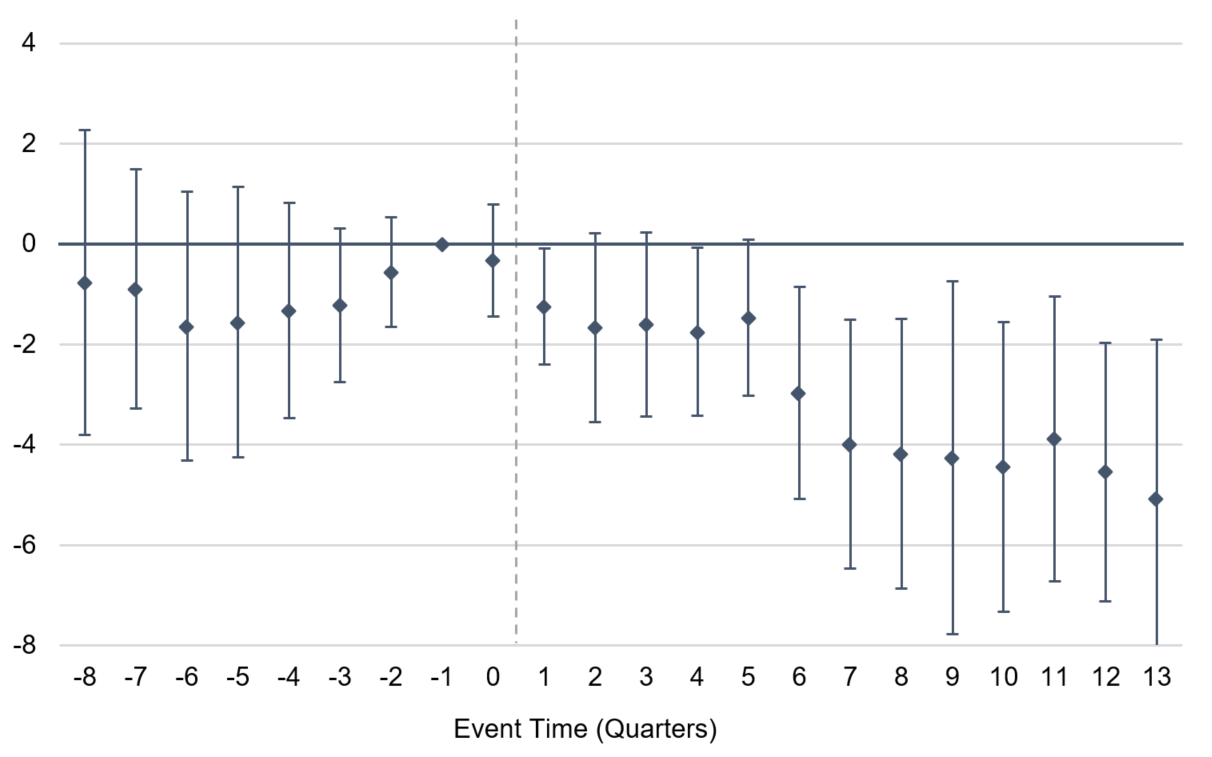} \\
B: Controlling for State-by-Calendar Quarter \\
\includegraphics[width=0.9\linewidth]{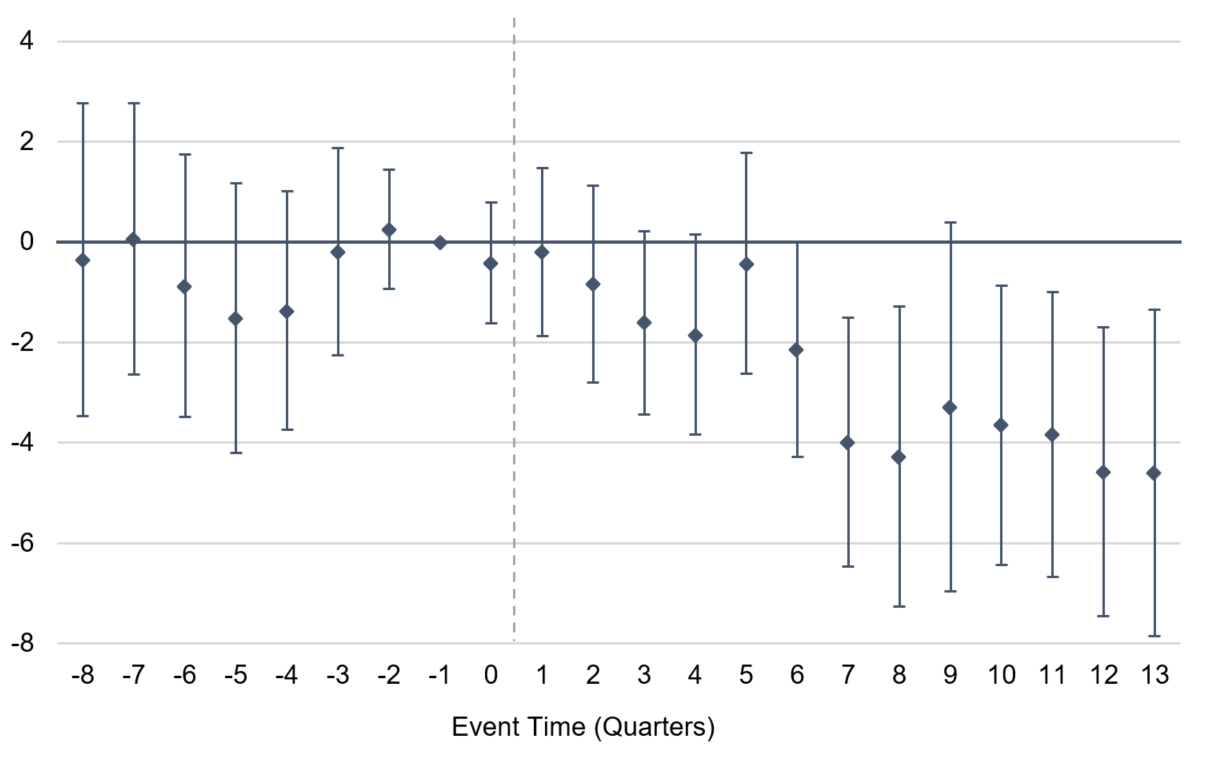} \\
\end{minipage}
\singlespacing
N = 651,000. Outcome variables denoted in percentage points. Also includes a control for state minimum wage, and includes SEIN and Year-Quarter-N2 fixed effects. Panel B also includes state-by-calendar quarter fixed effects. Error bars denote 95\% CI for robust standard errors clustered at state level.
\end{figure}

Turning next to the employment margin, Panel A of \autoref{growthyoy} plots baseline estimates of year-over-year employment growth, and Panel B includes state-by-calendar quarter fixed effects. Employment growth shows a sustained decline in treatment states after the tax increase, with a 2 percentage point drop by the end of the first year, and a 4 percentage point drop after the second. This growing treatment effect over time could be driven by two factors. Looking at a breakdown of taxes by treatment state in \autoref{statebystate2}, we see that some states experienced a gradual tax increase that peaked in 2012 (either a year or two after the initial increase). Secondly, average tax rates were also highest in 2011-2012, as layoffs from the Great Recession took a few years to translate into higher tax rates. Panel B shows that this pattern is also robust to the inclusion of state-specific calendar quarter fixed effects, which account for differences in geographic seasonality.

\subsection{Difference-in-Differences}
I next use a difference-in-differences estimation strategy to estimate average treatment magnitudes. In these specifications, I now use a continuous treatment measure equal to the change in the maximum UI tax schedule between time $t$ and 2009, and equal to zero prior to 2009. I estimate specifications described by \autoref{eq2}, which include employer (SEIN) and industry-by-time fixed effects, as well as a control for the state minimum wage.

\autoref{earntable} reports estimates for three earnings outcomes, with unweighted estimates in the odd columns, and estimates weighted by firm employment in the even columns. The baseline specifications are unweighted, since the analysis sample is already restricted to SEINs between 20 and 500 workers in 2009, although weighting generally has little impact on the estimated magnitudes. Columns 1 and 2 calculate the employer's average quarterly earnings across stable workers who were employed for the full quarter, while Columns 3 and 4 include all workers when calculating the average. Both outcomes show a small increase in log quarterly earnings after the tax change in all quarters except for Q1, consistent with the observed pre-trend in \autoref{earn}. The absence of an increase in the Q1 DD coefficient implies a maximum tax incidence of 40\% for every additional dollar of UI tax (given that \$100 is roughly 1\% of mean quarterly earnings), although this effect is likely driven by seasonal employment patterns and is not robust to a triple-difference specification.

In Columns 5 and 6 I estimate an alternative measure, which is how likely workers are to experience an earnings increase in a given quarter. Because quarterly earnings vary throughout the year (especially when inclusive of hours), I define a simple indicator for whether the worker earned more in the current quarter than in the previous two quarters (ignoring any workers with less than 3 quarters of tenure). The firm-level outcome is then calculated as the share of workers satisfying the previous definition. This measure is expected to be correlated with the share actually receiving raises, but estimates are likely to be attenuated by measurement error due to the quarterly frequency. These estimates provide further evidence for a minor degree of tax incidence in the first quarter of the year; for every \$100 increase in per-capita maximum UI taxes, the share of workers receiving a raise in earnings in Q1 falls by 0.4 percentage points.

Given the evidence against full tax incidence on earnings, I next investigate employment-related outcomes. \autoref{emptable} reports estimates for three different outcomes, and estimates are robust to weighting by firm employment (even columns). I define year-over-year employment growth as the DHS growth rate of quarterly employment relative to the same quarter of the previous year,  $\frac{Emp_{t}-Emp_{t-1}}{\frac{1}{2}(Emp_{t} + Emp_{t-1})} $, which creates a symmetric growth measure that allows for entry and exit. The baseline unweighted specification in Column 1 estimates that UI tax increases lower year-over-year employment growth, and the effect compounds throughout the year. A \$100 increase in per-capita UI taxes lowers employment growth by 0.43 percentage points by the end of the year, and the magnitude is even larger if weighted by firm employment. This implies that in a state like Indiana, which experienced a \$500 jump in maximum UI taxes in 2011, employment growth fell by 2.2 percentage points in high exposure firms, relative to similar counterparts in control states.

To decompose this employment change into hires and separations, I define the quarterly new-hire share as the number of new hires divided by employment, and the quarterly separation share as the number of separations divided by employment. The estimates show a negative impact on hiring, with no statistically significant effect on separations. Column 3 finds that a \$100 increase in per-capita maximum UI taxes lowers both the first quarter and last quarter hiring rate by 0.17 percentage points. This pattern of negative employment growth and hiring is consistent with the mechanism of greater UI tax burdens. At the start of the calendar year all earnings up to the tax base are eligible for UI taxes, and thus create a large payroll tax burden for the employer. Each additional worker hired incurs an additional UI tax burden, whereas additional separations do not incur a tax saving and could even push future costs higher if they claim UI benefits. Towards the end of the calendar year, most if not all existing workers have exceeded the UI tax base and no longer cost their employer any additional payroll tax. However, the earnings of new hires still incur taxes up to the same tax base, resulting in a much higher effective tax rate for new hires; this could explain the observed reluctance to hire in the last quarter of the year. Although I am unable to observe this, another possible explanation for the drop in labor demand is substitution to contract workers, who would not be subject to UI taxation.\footnote{For example, \cite{kugleretal} found that payroll tax cuts in Colombia increased rates of formal employment among affected workers.}

\subsection{Heterogeneity by Worker and Employer Characteristics} 
To allow for heterogeneous responses for different types of workers, \autoref{subgroup} analyzes employment outcomes for three specific groups of workers. First, I study the impact on the hiring rate of low-earning workers, who are defined to be those who earned less than \$3000 in the quarter after being hired (inflation adjusted to 2015 dollars).\footnote{The \$3000 threshold was chosen to coincide with the federal poverty level for a single person household (\$11,770 in 2015).} Column 1 shows that a \$100 increase in the maximum UI tax lowered the Q1 share of low-earning new hires by 0.1pp; benchmarked to the sample mean of 6.5\%, this indicates a larger negative response than for the overall share of new hires from the previous table. 

Next I calculate year-over-year employment growth rates for two subgroups of interest: workers younger than 25 years of age, and those with only a high school education or less (conditional on being 25 or older).\footnote{Outcomes for workers older than age 55 were also estimated, but the results did not warrant submission for disclosure.} Since small firms may not always employ these types of workers, growth rates are defined to equal zero if a firm did not employ any of these workers in either the current or previous calendar quarter, resulting in a larger difference between the weighted and unweighted estimates when weighting by firm employment. Recall the previous estimate of a \$100 increase in maximum UI taxes lowering employment growth in Q4 by 0.43 percentage points (Column 1 of \autoref{emptable}). For young workers, the impact is an even larger decline of 0.66 percentage points for each \$100 increase. Growth rates for low-educated workers are more similar to the overall effect, with an estimated decline of 0.39 percentage points by the end of the year. This suggests that young workers who are just entering the workforce are the most impacted by falls in labor demand due to UI tax increases.

To test for heterogeneous responses by different types of employers, I classify employers based on their characteristics at baseline (2009:Q3). I focus on three observable characteristics: median worker earnings, multi-establishment status, and firm age. The latter two are also potential proxies for whether the employer would face cash or financing constraints, as a surprise increase in UI taxes could restrict available cash flow. Recall that all firms in the analysis sample are already age 5 or older by 2010, as they needed a long enough layoff history to be identified as high exposure. 

\begin{figure}[H]
\begin{minipage}{\linewidth}
    \caption{Heterogeneity in New Hire Share by Firm Type (2008-2013)}
            \label{heterogeneity}
    \centering
\includegraphics[width=\linewidth]{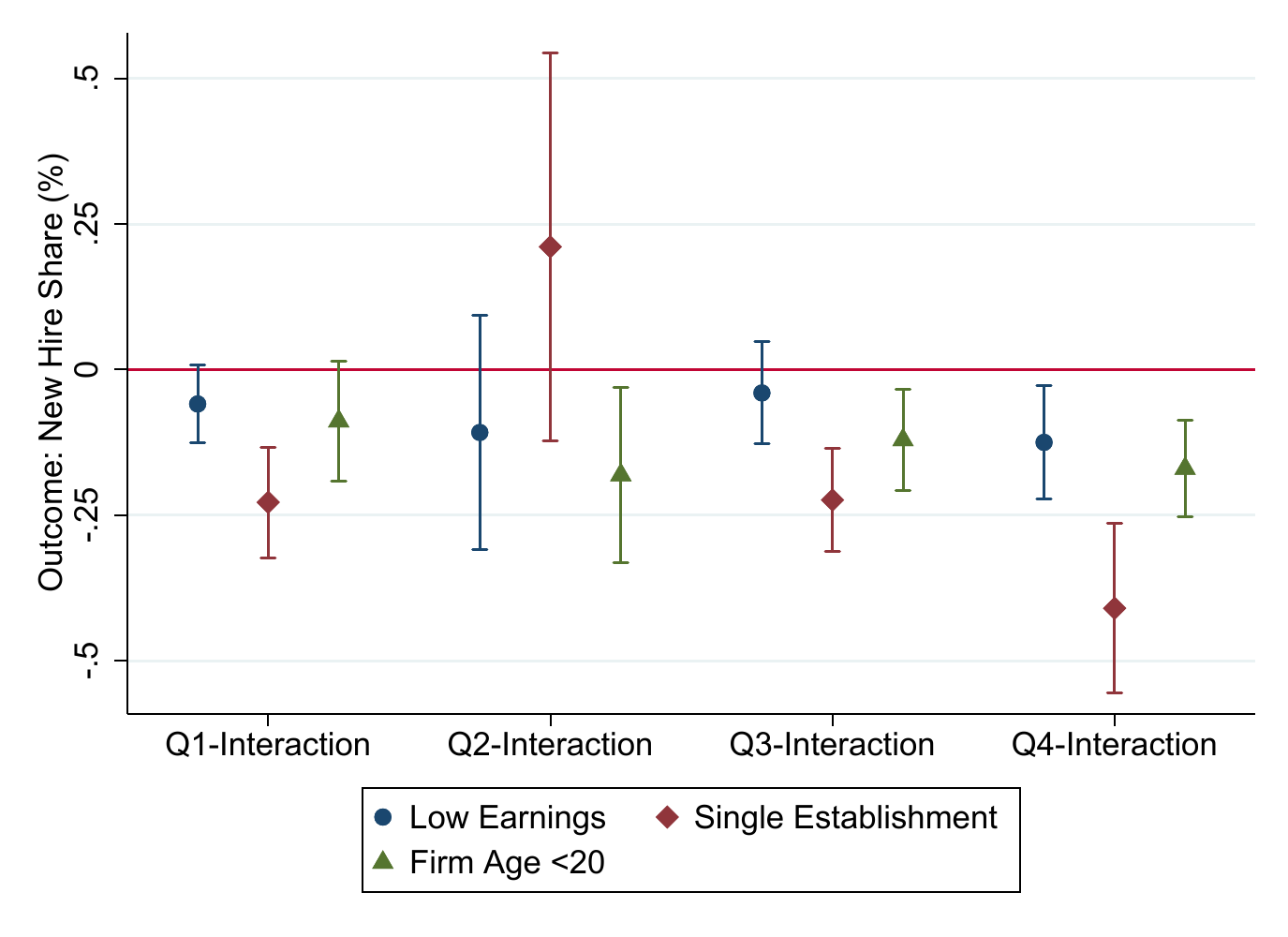}
\end{minipage}
\singlespacing
N = 651,000. Also includes a control for state minimum wage, and includes SEIN and Year-Quarter-N2 fixed effects. Error bars denote 95\% CI for robust standard errors clustered at state level. The share of employers with median earnings below \$6000 equals 0.29. The share that are single-establishment firms equals 0.69. The share with firm age under 20 equals 0.22.
\end{figure}

The main outcome of interest is the new hire share, and the estimation augments the differences-in-differences specification described by \autoref{eq2}. In each specification, I add four additional interaction terms interacting UI tax changes with one of the following indicators defined at baseline: Median quarterly earnings below \$6000, single-establishment firm (both in-state and nationally), or firm age less than 20 years old. \autoref{heterogeneity} plots estimates for each of the three characteristics. The estimates show little difference in hiring whether an employer has a majority of low-earning workers or not. However, the single establishment interactions show a large negative response for firms that only have a single location (seen by the interactions for Q1, Q3, and Q4), while multi-establishment firms are essentially unaffected. This indicates that single-establishment firms with less access to resources are the most likely to lower hiring in response to UI tax increases. Among these employers, a \$100 maximum tax hike lowered hiring by 0.23pp, 0.22pp, and 0.41pp in quarters 1, 3, and 4 respectively. Finally, the last set of markers shows that younger firms are also more likely than older firms to reduce hiring in response to a UI tax increase, although the magnitudes are smaller than those for single-establishment firms.

Appendix \autoref{het3} plots estimates for the separations share as well. While I estimated no overall effect on separations, the quarterly interactions suggest that single establishment firms exposed to a tax increase were actually less likely to experience separations in the first two quarters of the year, relative to multi-establishment firms. A potential mechanism could be a greater effort towards worker retention in the absence of hiring.

\subsection{Triple-Difference Design and Robustness}
In order to control for state-specific factors that could be confounding our estimates, I also conduct a difference-in-difference-in-differences estimation comparing the main sample of high-exposure employers to the placebo sample that experienced minimal layoffs. Because UI tax increases are more binding for employers near the maximum tax rate, we can use employers with minimal layoffs -- and therefore low exposure to UI tax increases -- to control for state-level variation that may influence both tax policy and employment outcomes.

\begin{figure}[H]
\begin{minipage}{\linewidth}
    \caption{Year-Over-Year Employment Growth - Triple Difference (2008-2013)}
            \label{growth_triple}
    \centering
\includegraphics[width=\linewidth]{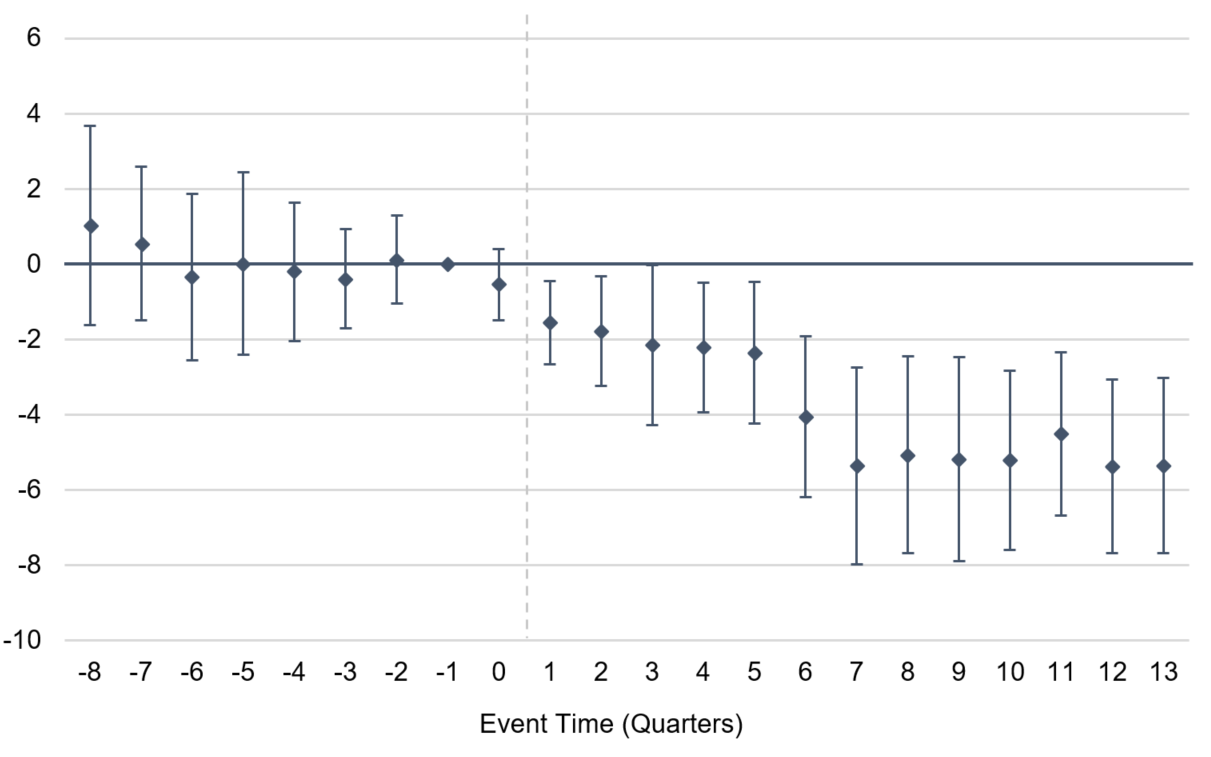}
\end{minipage}
\singlespacing
N = 1,109,000. Figure plots the estimates from treatment dummies interacted with event time and high-tax status. Regression also includes treatment interacted with event time, a control for state minimum wage, and SEIN and Year-Quarter-N2-High fixed effects. Error bars denote 95\% CI for robust standard errors clustered at state level.
\end{figure}

\autoref{growth_triple} plots event study estimates for year-over-year employment growth using this triple differences design on the expanded sample of both high tax and low tax employers. It estimates \autoref{eq1} with not only the event-time treatment dummies, but also a set of treatment dummies interacted with whether the employer is classified as high-tax. These interactions become our new coefficients of interest, while the baseline event-time estimates control for confounding conditions in treatment states. The estimates show no differential pre-trend prior to the tax increases, and a sustained drop in growth after, reaching negative 5 percentage points by the end of the second year. This provides further evidence of a causal relationship between UI tax increases and drops in labor demand.

\autoref{triplediff} reports estimates of a pooled triple-difference regression specification studying earnings, hiring, and separations. While the Taxchange*High coefficients for the new hire share and separation share are consistent with those from the baseline difference-in-differences specification, we no longer observe an impact on log earnings or the share receiving raises. This implies that the previously observed negative impact on Q1 earnings and raises was due to underlying differences in treatment and control states unrelated to UI tax increases, and the wage incidence of UI tax increases are minimal. 

\autoref{triplediff2} reports estimates for employment growth outcomes. The triple-difference interactions are all statistically significant and larger in magnitude than the original difference-in-differences estimates. Within just the first quarter of the year, employment growth drops by 0.5 percentage points for every \$100 UI tax increase. And this negative effect grows to -0.74pp by the end of the year. The employment growth of young workers were even more negatively impacted, with each \$100 in tax increases lowering end-of-year growth by 0.9pp. These results are also robust to the inclusion of state-by-year-quarter fixed effects, reported in \autoref{triplediff_styear}.

My results are also robust to a number of sensitivity checks of sample composition.\footnote{Estimates from robustness checks have not been disclosed from the Census RDC, but would be available for additional disclosure should they be required.} First, while I opted to exclude the retail trade and food/hospitality sectors due to higher likelihood for measurement error, both the difference-in-differences and triple-differences estimates are robust to the inclusion of these employers. Second, because I do not observe actual employer-specific tax rates, I measure expected UI tax costs based on imputed layoff histories. Small changes to the cutoff of layoff shares that warrant inclusion in my main analysis sample had no qualitative effect on my estimates. Additionally, the results are not driven by the inclusion of any one state in the analysis sample, nor does the restriction to a balanced panel impact the estimates. 

Finally, to address potential concerns with the differential timing of state UI tax increases, I have estimated separate event study regressions including only 2010 treatment states or only 2011 treatment states, respectively. This ensures that each group of treatment states are only compared to the control states, and there is no longer an issue of staggered treatment, since all tax increases occur at the same time. Treatment effects from this exercise are qualitatively similar to those using the pooled sample.

\section{Incidence for Low-Earning Workers}
Due to low UI tax bases (the median tax base in 2009 was only \$10,000), low-wage workers face higher effective tax rates and thus bear a larger relative tax burden than their higher-earning counterparts. They may also have less bargaining power and are more likely to work part-time. I identify existing workers of highly exposed firms, and estimate the impact of UI tax increases on their subsequent earnings trajectories. The added precision of a worker-level analysis allows for the identification of earnings impacts within this subgroup of interest.

Focusing on the main analysis sample of high-exposure firms, I construct a worker-level sample consisting of workers employed by the firm for at least a year, with annual earnings between \$5000 and \$24,000 in 2009.\footnote{Because the LEHD reports total earnings and not wage rates, this definition may pick up high-wage workers who work very few hours, although this should be fairly uncommon. Additionally, since UI taxes are based on earnings and not wages, part-time high-wage workers are equally costly to an employer from a payroll tax standpoint.} To ensure these workers make up a non-negligible share of the firms' employment, I also drop any firms that did not employ 10 or more of these low-wage workers as of 2009:Q3. This subsample includes approximately 549,000 low-wage workers, employed by 13,000 employers (44\% of the highly exposed employers in the main analysis sample). I estimate worker-level event study regressions, where event time is defined in the same way as \autoref{eq1}.
\begin{multline}
EarnGrowth_{ifst}  = \alpha_{f} + \sum_{k=-8, k\neq -1}^{8}\beta_{k}Taxchange_{s}(t - Increase_{s} = k) \\  + \delta_{t} +  \rho log(earn_{ifs,t-4}) + \gamma minwage_{st} + \epsilon_{ifst}
       \label{eq3}
\end{multline}

Here $i$ indexes worker, $f$ indexes employer, $s$ indexes state, $t$ indexes year-quarter, and $k$ indexes the quarter relative to the policy change. I group all quarters more than two years before the policy change together into one estimate $k=-8$, and I follow workers for up to three years after the policy change, through the end of 2012. $Taxchange_{s}$ is defined for treatment states to equal the average maximum UI tax increase between 2009 and 2010-11 (in hundreds), and equal to zero for control states. The main outcome of interest, $EarnGrowth_{ifst}$ is defined as the year-over-year earnings growth relative to the same quarter in the previous year. Thus inclusion in the sample requires the worker to have been employed at the firm a year earlier, limiting the analysis to a more stable subset of workers. 

\begin{figure}[H]
\begin{minipage}{\linewidth}
    \caption{Year-Over-Year Earnings Growth - Worker Level (2008-2012)}
            \label{workerlevel}
    \centering
\includegraphics[width=\linewidth]{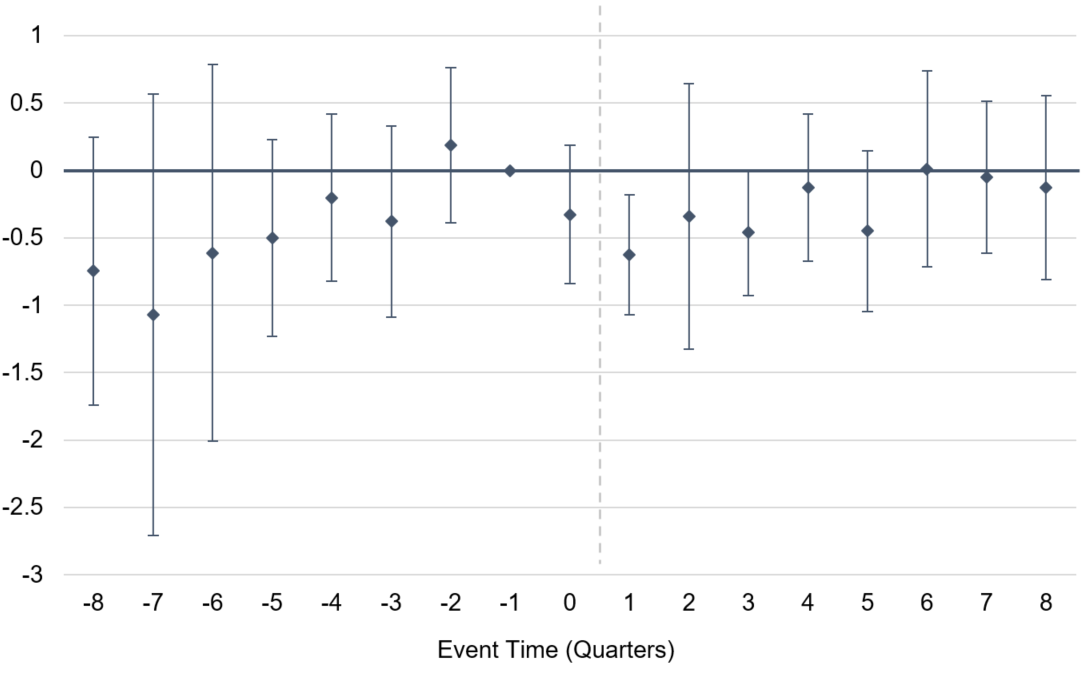}
\end{minipage}
\singlespacing
N = 4,954,000. Includes 549,000 unique workers across 13,000 unique employers. Figure plots the estimates from taxchange measure interacted with event time. Regression also includes a control for state minimum wage, lagged log earnings, and SEIN, Year-Quarter, and state-by-calendar quarter fixed effects. Error bars denote 95\% CI for robust standard errors clustered at state level.
\end{figure}

\autoref{workerlevel} plots the estimates of the event time interactions with $Taxchange_{s}$; thus the magnitude of each estimate reflects the effect of a \$100 UI tax increase. In the first quarter after the tax increase (event time = 1), each \$100 UI tax increase lowered earnings growth by 0.63\% for continuing workers in treated firms, relative to their counterparts in control firms. This negative impact is small and precisely estimated, and does not last beyond the first year. This allows us to rule out negative earnings impacts beyond 1\% in Q1 and 0.9\% in Q3, for each \$100 increase. Since 1\% of quarterly earnings equates to \$48 on average among this subsample, together these estimates imply a maximum pass-through rate of 90\% (\$90 out of \$100) for existing low-wage workers. Therefore, even amongst a subgroup that bears the largest effective tax burden, we find evidence that short-run tax increases result in less than full tax incidence on the earnings of existing workers.

\section{Conclusion}
The employer-specific and time-varying aspects of UI payroll taxation provides researchers with extensive policy variation that can be used to understand how firms respond to temporary payroll tax changes. Given previous proposals of cutting employer payroll taxes to stimulate employment during recessions, there is debate over whether such tax cuts would have the intended effect or whether they would simply increase firm profits. I estimate significant negative employment responses to unexpected tax increases, which suggests that temporary payroll tax cuts are likely to have a positive impact on employment as well, especially for young and low-earning workers that face higher relative tax burdens, and for smaller employers with less access to financial resources.

In response to state maximum UI tax increases in the years following the Great Recession, UI tax burdens rose for employers with large layoff histories, with minimal evidence of tax incidence on workers. With little ability to pass tax increases on to workers, employers lowered hiring rates and employment growth, and this effect grew over time. Using the Q1 treatment estimates of hiring rates from \autoref{emptable}, I calculate an implied labor demand elasticity of -1.1. This measure is obtained by dividing the Q1 DD coefficient estimate of -0.170 by 0.161, the first quarter effective UI tax increase estimated using the QCEW in Appendix \autoref{erateDD}. An alternative measure can be calculated by dividing the triple-difference drop in Q4 employment growth of -0.74 by 0.31, the share of average annual earnings a \$100 tax increase is equivalent to; this equates to an end-of-year labor demand elasticity of -2.4. These elasticity estimates fall within the range of elasticities from the labor demand literature, although they are on the lower end of those from recent payroll tax studies.\footnote{\cite{kuetal}, estimates a labor demand elasticity of -3.6 using place-based payroll tax increases in Norway, while \cite{benzarti2} estimates elasticities of -2.9 to -4.16 in response to firm-specific increases in Finland. And in the U.S. setting, \cite{johnston} estimates a UI tax labor demand elasticity of -4 while \cite{guo} estimates an elasticity of -1.1.}

The tax incidence of unemployment insurance payroll taxes has been an understudied topic in the payroll taxation literature. The experience rating of employer-specific tax rates creates the benefit of reducing avoidable layoffs (particularly temporary layoffs) and the associated externalities imposed upon the UI system. However, the cyclical pattern in which tax rates are highest in years following economic recessions also slows hiring and employment growth during the recovery. More work is needed to understand the unintended consequences and inherent tradeoffs of US unemployment insurance's current design.

\newpage
\onehalfspacing
\bibliography{references}

\newpage
\noindent \Large{\textbf{FIGURES}}

\newpage
\noindent \Large{\textbf{TABLES}}

\begin{table}[htbp]\centering
\def\sym#1{\ifmmode^{#1}\else\(^{#1}\)\fi}
\caption{Summary Statistics (2009:Q3)}
\begin{tabular}{l*{4}{cccc}}
\hline\hline
            &\multicolumn{2}{c}{Control}  &\multicolumn{2}{c}{Treatment} \\
            &       \phantom{00} Mean \phantom{00} &              \phantom{00} SD \phantom{00} &     \phantom{00} Mean \phantom{00} &        \phantom{00} SD \phantom{00}     \\
\hline \\
Employment &	66.74 &	72.28	 & 67.29 &	70 \\
[0.5em]
Cumulative Layoff Share 2006-07 &	0.5668	& 0.1689 &	0.5668 &	0.1679 \\
[0.5em]
Cumulative Layoff Share 2008-09 &	0.5037	& 0.1473 &	0.5065	& 0.1493 \\
[0.5em]
Average Annual Earnings &	36,900	& 27,990	& 32,350 &	23,340 \\ 
[0.5em]
Share of Workers Earning $<\$5000$	& 0.1518	& 0.1647 &	0.1572 &	0.1619 \\
[0.5em]
Construction Sector &	0.245 &  &	0.2282 &  \\
[0.5em]
\hline
\(N\)              &\multicolumn{2}{c}{18000}    &\multicolumn{2}{c}{11500}    \\
\hline\hline
\label{summstats}
\end{tabular}
  \begin{minipage}{15cm}
  \singlespacing
Employer-level observations from 3rd quarter of 2009. Control states include AZ, CA, CO, DE, DC, FL, NE, and MO. Treatment states include AR, IN, IL, ME, MD, OK, SC, TN, and WV.
  \end{minipage}
\end{table}

\begin{table}[htbp]\centering
\def\sym#1{\ifmmode^{#1}\else\(^{#1}\)\fi}
\caption{Log Quarterly Earnings (2008--2013)}
\begin{tabular}{l*{6}{c}}
\hline\hline
            &\multicolumn{2}{c}{Stable Workers}&\multicolumn{2}{c}{All Workers}&\multicolumn{2}{c}{Share with Raise}\\
            &\multicolumn{1}{c}{(1)}&\multicolumn{1}{c}{(2)}&\multicolumn{1}{c}{(3)}&\multicolumn{1}{c}{(4)}&\multicolumn{1}{c}{(5)}&\multicolumn{1}{c}{(6)}\\

\hline
Tax $\Delta$*Q1 (\$100's)   &  -0.00113	& -0.000855	& 0.000158 &	-0.0000151	& -0.419\sym{***} &	-0.398\sym{***}
\\
            &     (0.00190)	& (0.00123)	& (0.00109)	& (0.000735)	& (0.113)	& (0.0805)     \\
[1em]
Tax $\Delta$*Q2 (\$100's)  &  0.00380\sym{***} &	0.00331\sym{***} &	0.00209\sym{**} &	0.00214\sym{***} &	0.111 &	0.137
 \\
            &   (0.000938)	& (0.000754)	& (0.000869) &	(0.000706) &	(0.0929) &	(0.157)
    \\
[1em]
Tax $\Delta$*Q3 (\$100's)   &   0.00353\sym{***} &	0.00334\sym{***} &	0.00309\sym{***} &	0.00287\sym{***} &	-0.00947	& -0.0320  \\
            &    (0.00101) &	(0.00101)	& (0.000864) &	(0.000729) & (0.0985) &	(0.0922)
      \\
[1em]
Tax $\Delta$*Q4 (\$100's)   &    0.00494\sym{***} &	0.00432\sym{***} &	0.00484\sym{***} &	0.00471\sym{***} &	0.0281 &	-0.0218  \\
            &   (0.00130) &	(0.000902) &	(0.00105) &	(0.000824) &	(0.139) &	(0.132)     \\
[1em]
Minimum Wage &    0.0152\sym{***} & 0.0121\sym{***} &	0.0181\sym{***} &	0.0126\sym{***} &	0.170 &	0.260
  \\
       &  (0.00279)	& (0.00276)	& (0.00363) &	(0.00316) & (0.227) &	(0.229)    \\
\hline
\(R^{2}\)   &   0.869 &	0.905 &	0.884 &	0.921 &	0.191 &	0.193  \\
Mean of Dep Var       &  9.232 &	9.240	& 9.067 &	9.057 &	31.0 &	31.7  \\
Weighting       &   & YES	&   & YES &	  & YES  \\
\(N\)       &       651000    &       651000       &     651000    &       651000            &       651000    &       651000             \\
\hline\hline
\label{earntable}
\end{tabular}
  \begin{minipage}{15cm}
    \begin{footnotesize}
     Regressions include SEIN and year-quarter-N2 fixed effects. Share with Raise equals the percentage of continuing workers who received a raise relative to the previous two quarters. Even columns are weighted by employment. Robust standard errors clustered at state level in parentheses. \sym{**} \(p<0.05\), \sym{***} \(p<0.01\)
    \end{footnotesize}
  \end{minipage}
\end{table}
\begin{table}[htbp]\centering
\def\sym#1{\ifmmode^{#1}\else\(^{#1}\)\fi}
\caption{Employment Outcomes (2008--2013)}
\begin{tabular}{l*{6}{c}}
\hline\hline
            &\multicolumn{2}{c}{Employment Growth}&\multicolumn{2}{c}{New Hire Share}&\multicolumn{2}{c}{Separation Share}\\
            &\multicolumn{1}{c}{(1)}&\multicolumn{1}{c}{(2)}&\multicolumn{1}{c}{(3)}&\multicolumn{1}{c}{(4)}&\multicolumn{1}{c}{(5)}&\multicolumn{1}{c}{(6)}\\

\hline
Tax $\Delta$*Q1 (\$100's)  &  -0.142	&  -0.469\sym{**}   &  -0.170\sym{**} &	-0.163\sym{**} &	-0.0968	& -0.0350  \\
            &  (0.129) &	(0.177) & (0.0723) &	(0.0672) &	(0.0832) &	(0.0583)    \\
[1em]
Tax $\Delta$*Q2 (\$100's) & -0.287\sym{**} &	-0.571\sym{**} &  0.240 &	0.224 & -0.0746 & -0.0494 \\
            & (0.120) &	(0.205) & (0.193) &	(0.168) &	(0.0897) &	(0.0757)   \\
[1em]
Tax $\Delta$*Q3 (\$100's) &-0.318\sym{**} &	-0.674\sym{***}
 & -0.104 &	-0.0962 &	0.0514 &	0.102\sym{**} \\
            &  (0.115) &	(0.194) & (0.0752) &	(0.0756) &	(0.0652) &	(0.0410)   \\
[1em]
Tax $\Delta$*Q4 (\$100's) & -0.431\sym{***} &	-0.759\sym{***} & -0.170\sym{*} &	-0.178\sym{*}	& 0.0587 &	0.110  \\
            &  (0.127)	& (0.171) & (0.0954) &	(0.0879) &	(0.169)	 & (0.147) \\
[1em]
Minimum Wage &  -0.912	 & -0.792 & -0.330\sym{*} &	-0.196 &	0.107 &	0.190  \\
       & (0.696) &	(0.781) & (0.182) &	(0.201) &	(0.208) &	(0.200) \\
\hline
\(R^{2}\)  & 0.193	& 0.254 & 0.427	& 0.529 &	0.405 &	0.521 \\
Mean of Dep Variable  & -4.201	& 0.349     & 13.40	& 14.98	 & 14.74 &	15.87 \\
Weighting       &   & YES	&   & YES &	  & YES 	 \\
\(N\)       &       651000    &        651000            &       651000    &       651000        &       651000    &       651000             \\
\hline\hline
\label{emptable}
\end{tabular}
  \begin{minipage}{15cm}
    \begin{footnotesize}
    Regressions include SEIN and year-quarter-N2 fixed effects. Outcomes denoted in percentage points, and even columns are weighted by employment. Robust standard errors clustered at state level in parentheses. \sym{*} \(p<0.10\), \sym{**} \(p<0.05\), \sym{***} \(p<0.01\)
    \end{footnotesize}
  \end{minipage}
\end{table}
\begin{table}[htbp]\centering
\def\sym#1{\ifmmode^{#1}\else\(^{#1}\)\fi}
\caption{Employment Growth of Subgroups (2008--2013)}
\begin{tabular}{l*{6}{c}}
\hline\hline
            &\multicolumn{2}{c}{}&\multicolumn{4}{c}{Employment Growth:}\\
            &\multicolumn{2}{c}{Low-Earn Hire Share}&\multicolumn{2}{c}{Age Under 25}&\multicolumn{2}{c}{HS or Less}\\
            &\multicolumn{1}{c}{(1)}&\multicolumn{1}{c}{(2)}&\multicolumn{1}{c}{(3)}&\multicolumn{1}{c}{(4)}&\multicolumn{1}{c}{(5)}&\multicolumn{1}{c}{(6)}\\

\hline
Tax $\Delta$*Q1 (\$100's) &  -0.102\sym{**} &	-0.0956\sym{**} &	-0.218 &	-0.511\sym{**} &	-0.0956 &	-0.450\sym{**}  \\
            &   (0.0460) &	(0.0433) &	(0.246) &	(0.234) &	(0.133) &	(0.182) \\
[1em]
Tax $\Delta$*Q2 (\$100's) & 0.0929	& 0.106 &	-0.325 &	-0.643\sym{**} &	-0.249\sym{*} &	-0.563\sym{**} \\
            &  (0.0777)	& (0.0694) &	(0.254) &	(0.226)	 & (0.122)	 & (0.209)	 \\
[1em]
Tax $\Delta$*Q3 (\$100's*) & -0.0701	 & -0.0457 &	-0.510\sym{*} &	-0.896\sym{***} &	-0.288\sym{**} &	-0.655\sym{***} \\
            &  (0.0521) &	(0.0434) &	(0.287) &	(0.250) &	(0.108) &	(0.200)  \\
[1em]
Tax $\Delta$*Q4 (\$100's) & -0.0822 &	-0.0714 &	-0.655\sym{*} &	-0.951\sym{***} &	-0.393\sym{***} &	-0.735\sym{***} \\
            &  (0.0543) &	(0.0454) &	(0.328) &	(0.264) &	(0.132) &	(0.172) \\
[1em]
Minimum Wage &  -0.237\sym{*} &	-0.151 &	0.0461 &	-0.446 &	-1.589\sym{*} &	-1.217  \\
       & (0.120) &	(0.118) &	(1.383) &	(1.497) &	(0.778) &	(0.836)   \\
\hline
\(R^{2}\)   & 0.515	& 0.647 &	0.101 &	0.154 &	0.156 &	0.218 \\
Mean of Dep Var       & 6.478 &	7.399 &	-11.44 &	-5.683 &	-2.712 &	1.972 \\
Weighting       &   & YES	&   & YES &	  & YES 	 \\
\(N\)          &     651000    &       651000            &       651000    &       651000 &       651000    &       651000          \\
\hline\hline
\label{subgroup}
\end{tabular}
  \begin{minipage}{15cm}
    \begin{footnotesize}
    Regressions include SEIN and year-quarter-N2 fixed effects. Outcomes are denoted in percentage points, and even columns are weighted by employment.  Robust standard errors clustered at state level in parentheses. \sym{*} \(p<0.10\), \sym{**} \(p<0.05\), \sym{***} \(p<0.01\)
    \end{footnotesize}
  \end{minipage}
\end{table}

\begin{table}[htbp]\centering
\def\sym#1{\ifmmode^{#1}\else\(^{#1}\)\fi}
\caption{Triple Difference (2008--2013)}
%\scalebox{0.9}{
\begin{tabular}{l*{4}{c}}
\hline\hline
            &\multicolumn{1}{c}{Log Earnings}&\multicolumn{1}{c}{Share with Raise}&\multicolumn{1}{c}{New Hire Share }&\multicolumn{1}{c}{Separation Share}\\          
            &\multicolumn{1}{c}{(1)}&\multicolumn{1}{c}{(2)}&\multicolumn{1}{c}{(3)}&\multicolumn{1}{c}{(4)}\\

\hline
Tax $\Delta$*Q1 (\$100's)   &  -0.000610 &	-0.247\sym{**} &	-0.0286\sym{*} &		-0.0385  \\
            &   (0.00147) &	(0.108) &	(0.0143) &	(0.0373) \\
[1em]
Tax $\Delta$*Q2 (\$100's) &  0.00313 &	0.130 &	0.0739 &	0.0305  \\
            &   (0.00274) &	(0.489) &	(0.0660) &		(0.0624) \\
[1em]
Tax $\Delta$*Q3 (\$100's) & -0.000553 &	0.0416 &	0.0382 &	-0.0480  \\
            &   (0.00283) &	(0.219) &	(0.0289) &		(0.0485) \\
[1em]
Tax $\Delta$*Q4 (\$100's) &  0.00291\sym{**} &	-0.000309 &	0.0380 &	-0.0719\sym{***}  \\
            &  (0.00128) &	(0.149) &	(0.0231) &		(0.0240) \\
[1em]
Tax $\Delta$*Q1 $\times$ High & -0.000246 &	-0.171 &	-0.155\sym{**} &	-0.0544  \\
            &  (0.00172) &	(0.114) &	(0.0654) &		(0.0508) \\
[1em]
Tax $\Delta$*Q2 $\times$ High & 0.00112 &	0.0285 &	0.155 &		-0.0995  \\
            &   (0.00261) &	(0.506) &	(0.156)	& 	(0.0574) \\
[1em]
Tax $\Delta$*Q3 $\times$ High &  0.00433 &	-0.0328 &	-0.151\sym{**} &		0.110  \\
            &   (0.00326) &	(0.278) &	(0.0596) &		(0.0797) \\
[1em]
Tax $\Delta$*Q4 $\times$ High  &  0.00223 &	0.0103 &	-0.220\sym{**} &	 0.135  \\
            &   (0.00172) &	(0.0873) &	(0.101) &		(0.164) \\
[1em]
Minimum Wage &  0.0108\sym{***} &	0.00649 &	-0.170 &		-0.00830 \\
            &  (0.00271) &	(0.262) &	(0.118) &		(0.144) \\
\hline
\(R^{2}\)   &  0.901 &	0.178 &	0.476 &	0.482 \\
Mean of Dep Var       & 9.323 &	31.7 &	10.61 &	11.28  \\
\(N\)       & 1109000 &	1109000 &	1109000 &	1109000       \\
\hline\hline
\label{triplediff}
\end{tabular}
  \begin{minipage}{15cm}
    \begin{footnotesize}
    Regressions include SEIN, year-quarter-N2, and year-quarter-High fixed effects. The necessary DDD interactions are either absorbed by the SEIN fixed effects or the year-quarter-High fixed effects. Robust standard errors clustered at state level in parentheses. \sym{*} \(p<0.10\), \sym{**} \(p<0.05\), \sym{***} \(p<0.01\)
    \end{footnotesize}
  \end{minipage}
\end{table}
\begin{table}[htbp]\centering
\def\sym#1{\ifmmode^{#1}\else\(^{#1}\)\fi}
\caption{Triple Difference - Employment Growth (2008--2013)}
\begin{tabular}{l*{3}{c}}
\hline\hline
            &\multicolumn{1}{c}{Overall Growth}&\multicolumn{1}{c}{Age under 25}&\multicolumn{1}{c}{HS or Less}\\          
            &\multicolumn{1}{c}{(1)}&\multicolumn{1}{c}{(2)}&\multicolumn{1}{c}{(3)}\\
\hline
Tax $\Delta$*Q1 (\$100's)   & 0.330\sym{**} &	0.379 &	0.371\sym{**}  \\
            &   (0.116)	& (0.228) &	(0.135) \\
[1em]
Tax $\Delta$*Q2 (\$100's) &  0.303\sym{**} & 	0.295 &	0.309\sym{*} \\
            & (0.127) &	(0.287) &	(0.154) \\
[1em]
Tax $\Delta$*Q3 (\$100's) &0.315\sym{**} &	0.208 &	0.357\sym{**} \\
            & (0.135) &	(0.242) &	(0.149) \\
[1em]
Tax $\Delta$*Q4 (\$100's) & 0.284\sym{*} &	0.222 &	0.299\sym{**} \\
            & (0.140) &	(0.265) &	(0.139) \\
[1em]
Tax $\Delta$*Q1 $\times$ High  & -0.503\sym{**}	& -0.618\sym{*}	& -0.525\sym{**}  \\
            &   (0.190)	& (0.293)	& (0.233) \\
[1em]
Tax $\Delta$*Q2 $\times$ High &-0.622\sym{***} &	-0.659\sym{**} &	-0.617\sym{**} \\
            &  (0.212)	& (0.290) &	(0.274) \\
[1em]
Tax $\Delta$*Q3 $\times$ High & -0.664\sym{***} &	-0.750\sym{**} &	-0.705\sym{**}  \\
            & (0.222) &	(0.326) &	(0.253)\\
[1em]
Tax $\Delta$*Q4 $\times$ High  & -0.743\sym{***} &	-0.904\sym{**} &	-0.748\sym{***}  \\
            &   (0.234) &	(0.387) &	(0.252) \\
[1em]
Minimum Wage &  -0.383 &	0.598 &	-0.801**  \\
            &  (0.336) &	(0.872)	 & (0.369) \\
\hline
\(R^{2}\)   &  0.203	& 0.086	& 0.167 \\
Mean of Dep Variable       & -1.373	& -7.296 &	0.419 \\
\(N\)       &   1109000	& 1109000 &		1109000       \\
\hline\hline
\label{triplediff2}
\end{tabular}
  \begin{minipage}{15cm}
    \begin{footnotesize}
    Growth measures are multiplied by 100. Regressions include SEIN, year-quarter-N2, and year-quarter-High fixed effects. The necessary DDD interactions are either absorbed by the SEIN fixed effects or the year-quarter-High fixed effects. Robust standard errors clustered at state level in parentheses. \sym{*} \(p<0.10\), \sym{**} \(p<0.05\), \sym{***} \(p<0.01\)
    \end{footnotesize}
  \end{minipage}
\end{table}

\appendix
\newpage
\section{APPENDIX TABLES AND FIGURES}
%\noindent \Large{\textbf{APPENDIX TABLES AND FIGURES}}

\renewcommand\thefigure{A.1}
\begin{figure}[H]
\caption{Empirical UI Tax Schedule for Florida (2008)}
\label{sample_sched}
\centering
\begin{minipage}{0.9\linewidth}
\includegraphics[width=\linewidth]{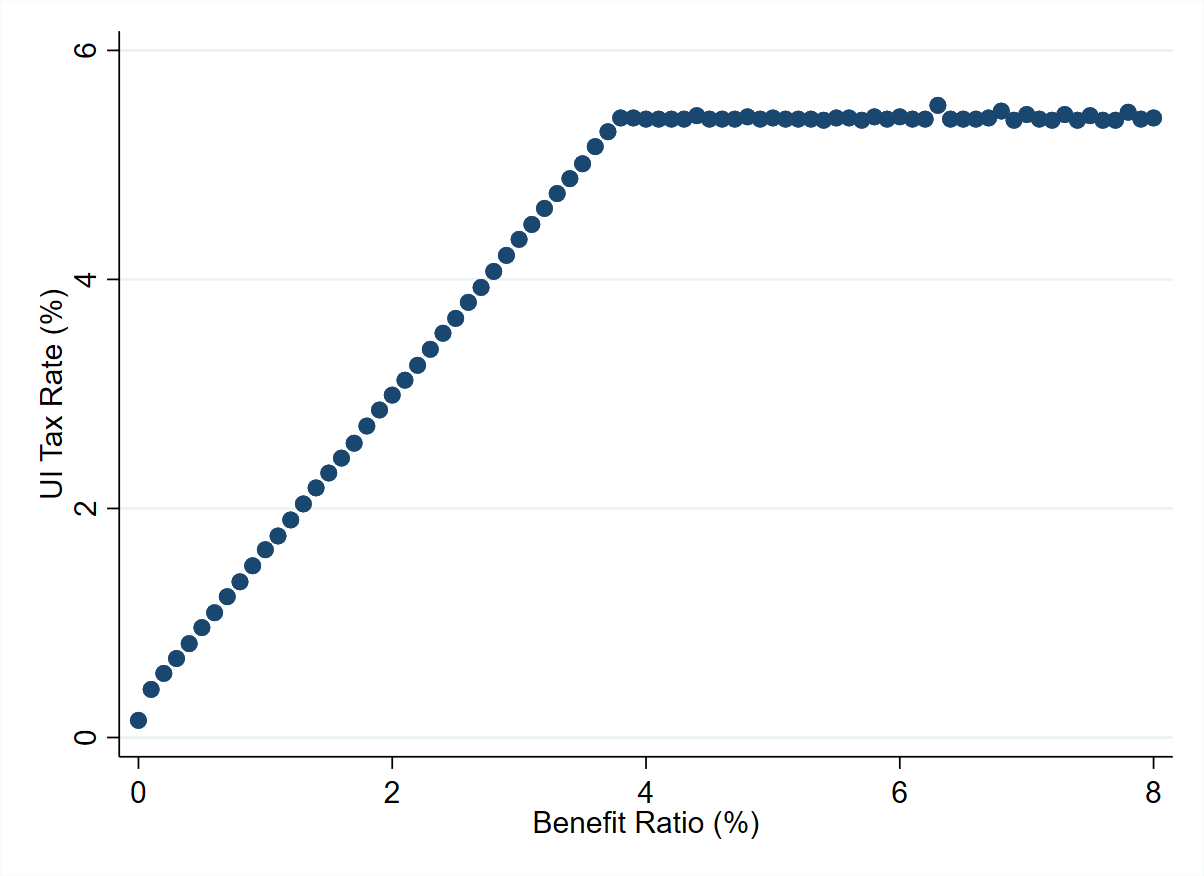}
   \footnotesize Source: US Dept of Labor ETA 204 Experience Rating Report. The Benefit Ratio is a measure of the
employer’s layoff experience in the last three years.
   \end{minipage}
\end{figure}

\renewcommand\thefigure{A.2}
\begin{figure}[H]
\caption{Total UI Contributions, as \% of total wages (1967 - 2018)}
\label{cycles}
\centering
\begin{minipage}{0.9\linewidth}
\includegraphics[width=\linewidth]{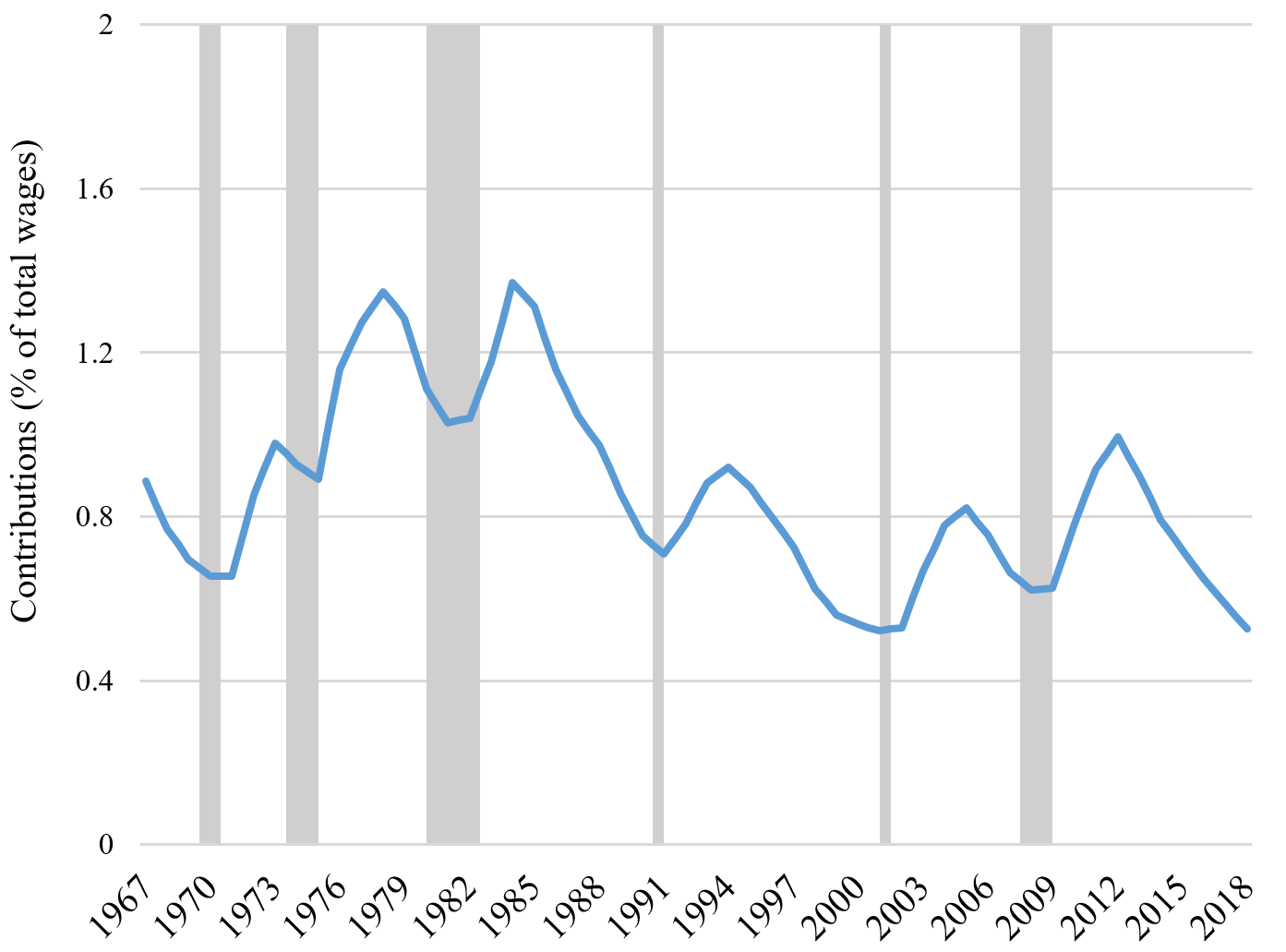}
   \footnotesize Source: US Dept of Labor Unemployment Insurance Chartbook. \\Shaded regions denote US recession years, as defined by NBER.
   \end{minipage}
\end{figure}

\renewcommand\thefigure{A.3}
\begin{figure}[H]
\caption{State UI Tax Schedules - Control}
\label{statebystate}
\centering
\begin{minipage}{0.9\linewidth}
\includegraphics[width=\linewidth]{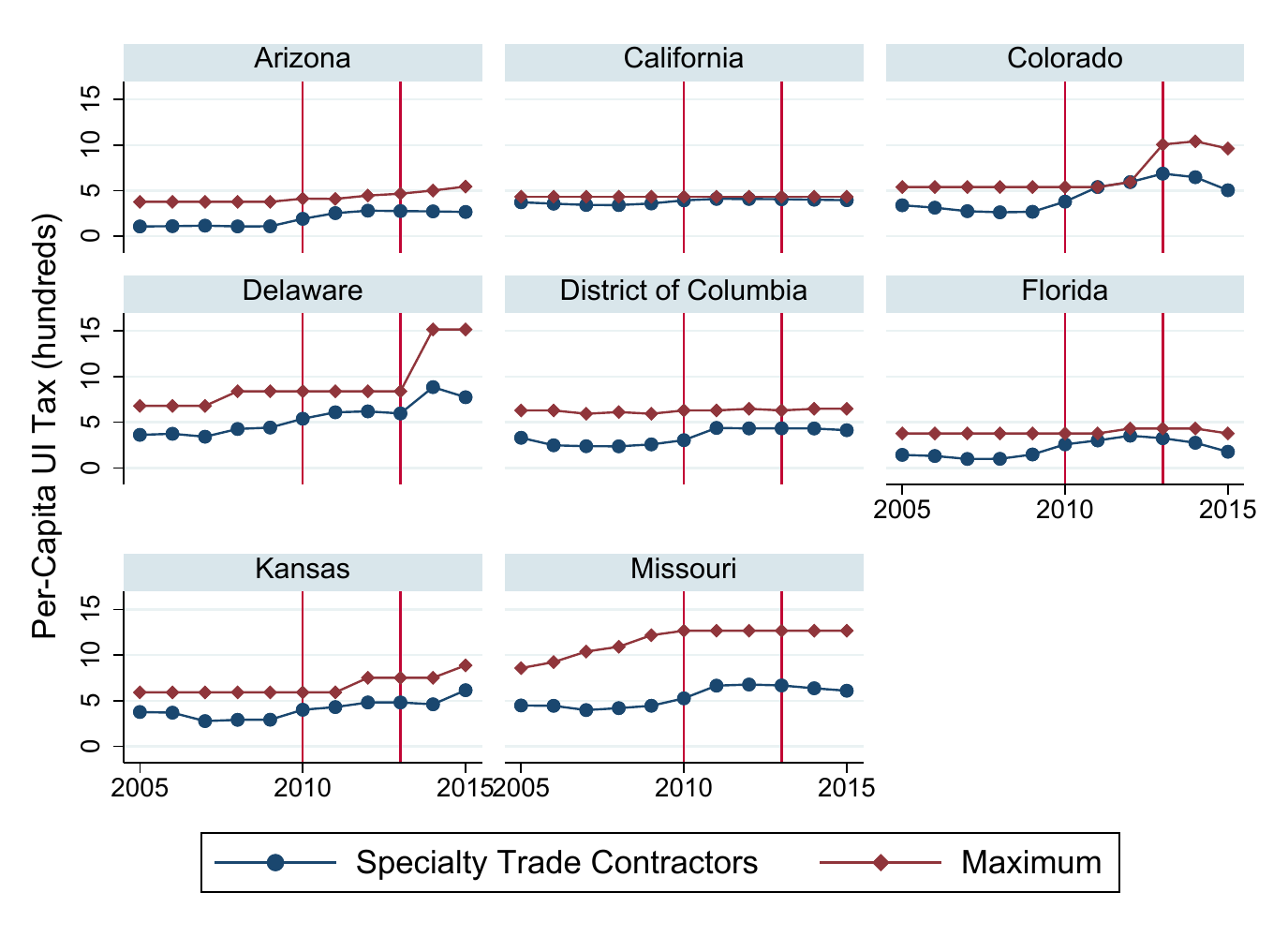}
   \footnotesize Compares average UI taxes paid and maximum UI taxes. Average taxes for specialty trade contractors obtained from Quarterly Census of Employment and Wages. Specialty trade contractors (NAICS 238) was chosen as an industry whose employers often face rates close to the maximum, to proxy for high exposure firms.
   \end{minipage}
\end{figure}

\renewcommand\thefigure{A.4}
\begin{figure}[H]
\caption{State UI Tax Schedules - Treatment}
 \label{statebystate2}
\centering
\begin{minipage}{0.9\linewidth}
\includegraphics[width=\linewidth]{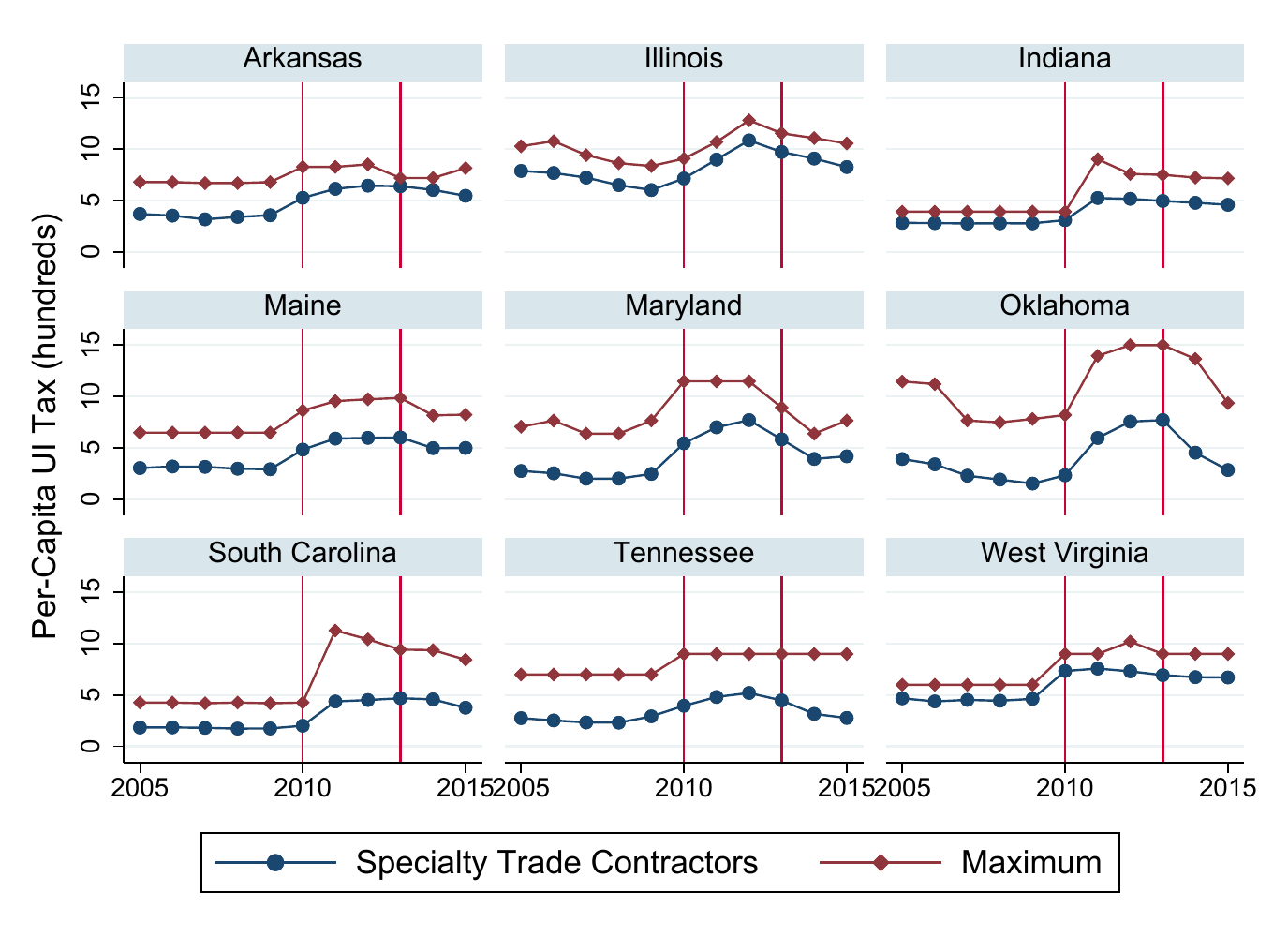}
   \footnotesize Compares average UI taxes paid and maximum UI taxes. Average taxes for specialty trade contractors obtained from Quarterly Census of Employment and Wages. Specialty trade contractors (NAICS 238) was chosen as an industry whose employers often face rates close to the maximum, to proxy for high exposure firms.
   \end{minipage}
\end{figure}

\renewcommand\thefigure{A.5}
\begin{figure}[H]
\caption{State UI Tax Schedules - Excluded}
 \label{statebystate3}
\centering
\begin{minipage}{0.9\linewidth}
\includegraphics[width=\linewidth]{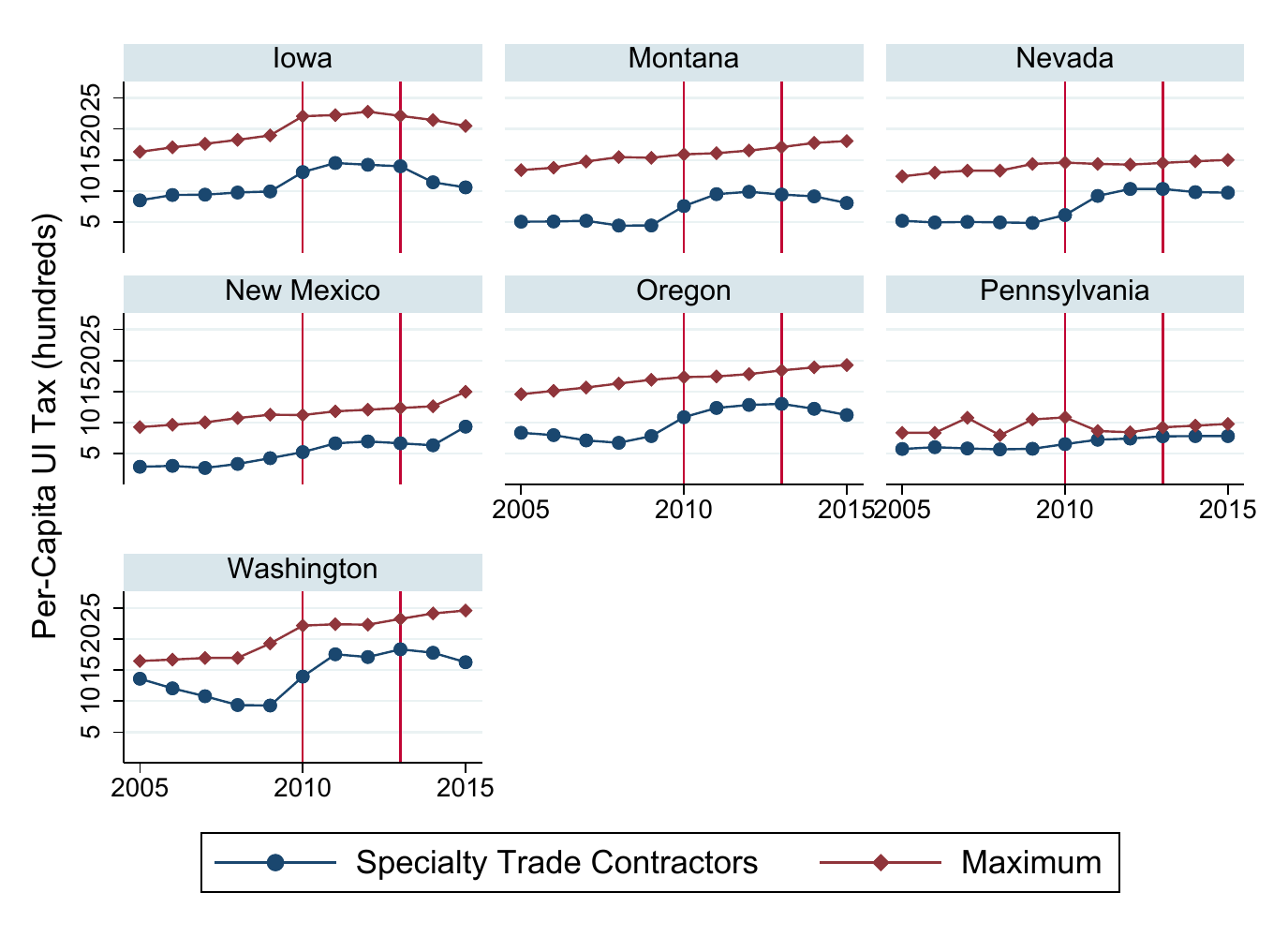}
   \footnotesize Compares average UI taxes paid and maximum UI taxes. Average taxes for specialty trade contractors obtained from Quarterly Census of Employment and Wages. Specialty trade contractors (NAICS 238) was chosen as an industry whose employers often face rates close to the maximum, to proxy for high exposure firms.
   \end{minipage}
\end{figure}

\renewcommand\thefigure{A.6}
\begin{figure}[H]
\begin{minipage}{\linewidth}
    \caption{Heterogeneity in Separations Share by Firm Type (2008-2013)}
            \label{het3}
    \centering
\includegraphics[width=\linewidth]{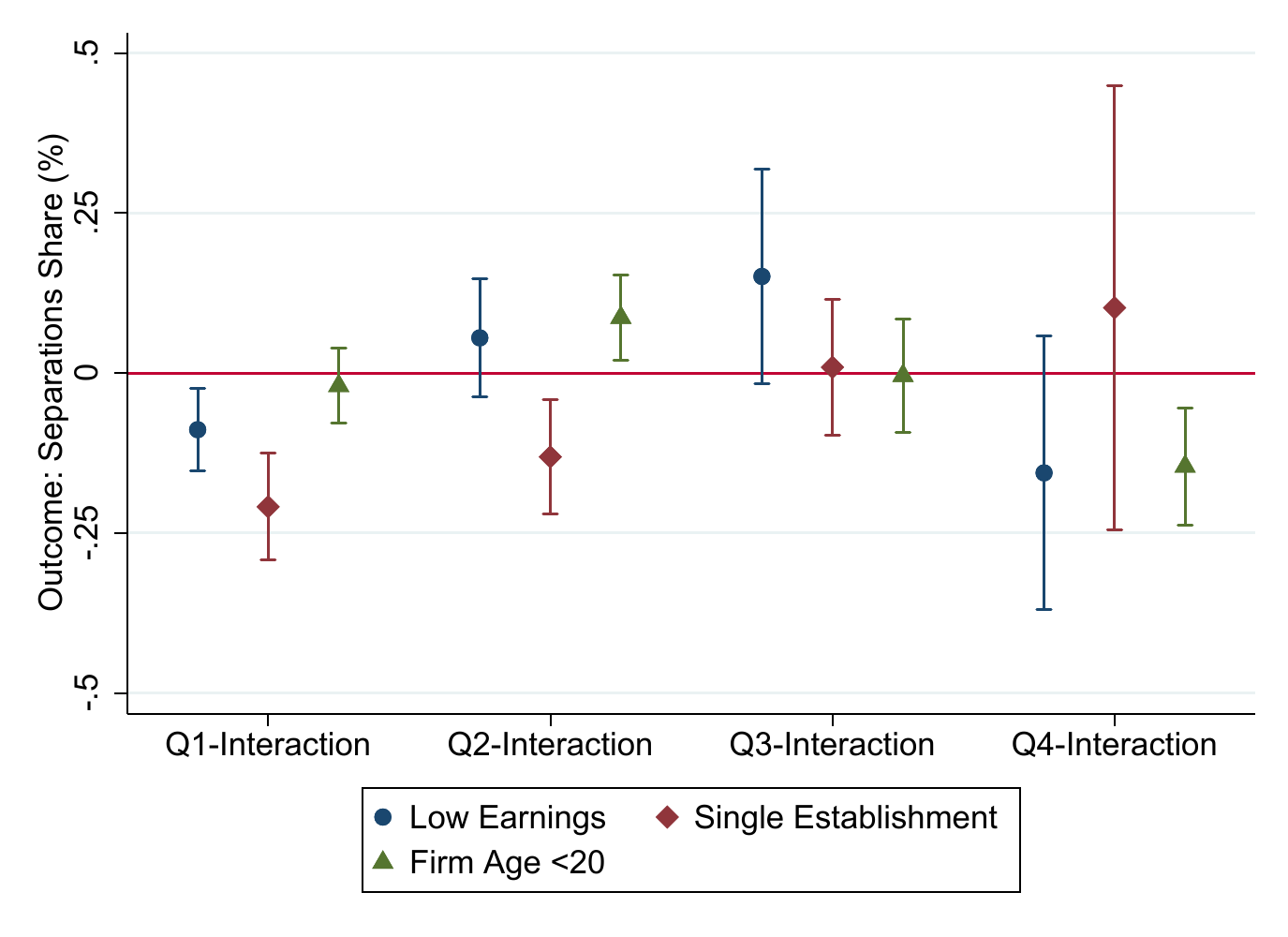}
\end{minipage}
\singlespacing
N = 651,000. Also includes a control for state minimum wage, and includes SEIN and Year-Quarter-N2 fixed effects. Error bars denote 95\% CI for robust standard errors clustered at state level. The share of employers with median earnings below \$6000 equals 0.29. The share that are single-establishment firms equals 0.69. The share with firm age under 20 equals 0.22.
\end{figure}

\renewcommand\thetable{A.1}
\begin{table}[htbp]\centering
\def\sym#1{\ifmmode^{#1}\else\(^{#1}\)\fi}
\caption{Average Industry Effective UI Tax Rate (2008--2013)}
\begin{tabular}{l*{2}{c}}
\hline\hline
            &\multicolumn{1}{c}{Unweighted}&\multicolumn{1}{c}{Employment-Weighted}\\
            &\multicolumn{1}{c}{(1)}&\multicolumn{1}{c}{(2)}\\

\hline
Tax $\Delta$*Q1 (\$100's)   & 0.0996\sym{*} &	0.161\sym{***}  \\
            &    (0.0498) &	(0.0646)   \\
[1em]
Tax $\Delta$*Q2 (\$100's) &  0.138\sym{***} &	0.136\sym{***}   \\
            &  (0.0242) &	(0.0270)    \\
[1em]
Tax $\Delta$*Q3 (\$100's) & 0.0752\sym{***} &	0.0490\sym{*} \\
            &   (0.0209)	& (0.0254)     \\
[1em]
Tax $\Delta$*Q4 (\$100's) &  0.0536\sym{*} & 0.0207 \\
            &  (0.0280) &	(0.0335)  \\
[1em]
Minimum Wage &  0.0865 &	0.206  \\
       & (0.104) &	(0.0987)    \\
\hline
\(R^{2}\)   &  0.881 &	0.914  \\
Mean of Dependent Variable       & 1.382	& 1.217  \\
\(N\)       &       4062    &       4062             \\
\hline\hline
\label{erateDD}
\end{tabular}
  \begin{minipage}{12cm}
    \begin{footnotesize}
    Uses average industry data from the public-use QCEW. Sample limited to ten 4-digit construction industries, and observations are at the industry by state by quarter level (cells too small to meet disclosure requirements are missing). Effective tax rates calculated by dividing quarterly UI contributions by quarterly payroll. Regressions include state-industry and year-quarter fixed effects. Robust standard errors clustered at state level in parentheses. \sym{**} \(p<0.05\), \sym{***} \(p<0.01\)
    \end{footnotesize}
  \end{minipage}
\end{table}

\renewcommand\thetable{A.2}
\begin{table}[htbp]\centering
\def\sym#1{\ifmmode^{#1}\else\(^{#1}\)\fi}
\caption{Triple Difference Controlling for State-Time FEs (2008--2013)}
\begin{tabular}{l*{6}{c}}
\hline\hline
            &\multicolumn{1}{c}{}&\multicolumn{1}{c}{}&\multicolumn{1}{c}{}&\multicolumn{1}{c}{}&\multicolumn{1}{c}{Growth:}&\multicolumn{1}{c}{Growth:}\\   
            &\multicolumn{1}{c}{Log Earn}&\multicolumn{1}{c}{Growth}&\multicolumn{1}{c}{New Hires}&\multicolumn{1}{c}{Separations}&\multicolumn{1}{c}{Under 25 }&\multicolumn{1}{c}{HS or Less}\\          
            &\multicolumn{1}{c}{(1)}&\multicolumn{1}{c}{(2)}&\multicolumn{1}{c}{(3)}&\multicolumn{1}{c}{(4)}&\multicolumn{1}{c}{(5)}&\multicolumn{1}{c}{(6)}\\

\hline
Tax $\Delta$*Q1 $\times$ High & -0.00292 &	-0.528\sym{**} &	-0.154\sym{*} &	-0.0456 &	-0.665\sym{**} 	& -0.551\sym{**}  \\
            &  (0.00320) & (0.190) &	(0.0739) &	(0.0601) &		(0.284) &	(0.231) \\
[1em]
Tax $\Delta$*Q2 $\times$ High & 0.00184 &	-0.642\sym{***} &	0.196 &	-0.0838 &	-0.717\sym{**} &	-0.640\sym{**} \\
            & (0.00322) &	(0.215) &	(0.185) &	(0.0528) &	(0.277) &	(0.275) \\
[1em]
Tax $\Delta$*Q3 $\times$ High &  0.00511\sym{**} &	-0.672\sym{***} &	-0.131\sym{**} &	0.130\sym{**} &	-0.776\sym{**} &	-0.723\sym{**}  \\
            &   (0.00239) &	(0.218) &	(0.0504) &	(0.0607) &	(0.301) &	(0.249) \\
[1em]
Tax $\Delta$*Q4 $\times$ High  & 0.00256 &	-0.737\sym{***} &  0.199\sym{*} &	0.195	& -0.939\sym{**} &	-0.754\sym{***}\\
            &   (0.00219) &	(0.227) &	(0.102)	& (0.181)	& (0.370) &	(0.244) \\
\hline
\(R^{2}\)   &  0.914 &	0.205 &	0.479 &	0.485 &	0.087 &	0.168 \\
Mean of Dep Var       & 9.096	& -1.373 &	10.61 &	11.28 &	-7.296	& 0.419  \\
\(N\)       & 1109000 &	1109000 & 1109000 &	1109000 &	1109000 &	1109000       \\
\hline\hline
\label{triplediff_styear}
        \end{tabular}
  \begin{minipage}{15cm}
    \begin{footnotesize}
    Outcomes in percentage points. Regressions include SEIN, year-quarter-state, year-quarter-N2, and year-quarter-High fixed effects. The necessary DDD interactions are either absorbed by the SEIN, year-quarter-state, or year-quarter-High fixed effects. Robust standard errors clustered at state level in parentheses. \sym{*} \(p<0.10\), \sym{**} \(p<0.05\), \sym{***} \(p<0.01\)
    \end{footnotesize}
  \end{minipage}
\end{table}

\end{document}